\documentclass[traditabstract,onecolumn]{aa}

\usepackage{graphicx}
\usepackage{natbib}
\usepackage{amssymb}

\newcommand{\bpm}{B_m^p(\omega)}
\newcommand{\bpam}{B^p_{a,m}(\omega)}
\newcommand{\etal}{{\it et al.}}
\newcommand{\aass}{{\it Astron. Astrophys. Suppl. Ser.}}

\newcommand{\colhead}{}

\begin{document}

\title{Circular polarization measurement in millimeter-wavelength
spectral-line VLBI observations}

\author{A. J. Kemball
        \inst{1,2}
        \and
        L. Richter\inst{2}}

\institute{Department of Astronomy and 
  Institute for Advanced Computing  Applications and Technologies,\\ 
  University of Illinois at Urbana-Champaign,\\ 
  1002 W. Green Street, Urbana, IL 61801, USA
  \and
  Department of Physics and Electronics,\\ 
  Rhodes University,\\ P.O. Box 94, Grahamstown, 6140, South Africa}


\abstract{
This paper considers the problem of accurate measurement of circular polarization in imaging spectral-line VLBI observations in the $\lambda=7$ mm and $\lambda=3$ mm wavelength bands. This capability is especially valuable for the full observational study of compact, polarized SiO\ maser components in the near-circumstellar environment of late-type, evolved stars. Circular VLBI\ polarimetry provides important constraints on SiO maser astrophysics, including the theory of polarized maser emission transport, and on the strength and distribution of the stellar magnetic field and its dynamical role in this critical circumstellar region. We perform an analysis here of the data model containing the instrumental factors that limit the accuracy of circular polarization measurements in such observations, and present a corresponding data reduction algorithm for their correction. The algorithm is an enhancement of existing spectral line VLBI polarimetry methods using autocorrelation data for calibration, but with innovations in bandpass determination, autocorrelation polarization self-calibration, and general optimizations for the case of low SNR, as applicable at these wavelengths. We present an example data reduction at $\lambda=7$ mm and derive an estimate of the predicted accuracy of the method of $m_c \leq 0.5\%$ or better at $\lambda=7$ mm and $m_c \leq 0.5-1\%$ or better at $\lambda=3$ mm. Both the strengths and weaknesses of the proposed algorithm are discussed, along with suggestions for future work.}


\titlerunning{Circular polarization mm-VLBI}
\authorrunning{Kemball and Richter}

\maketitle


\section{Introduction}

SiO maser emission acts as an important tracer of the
astrophysical conditions in the near-circumstellar environments of
late-type, evolved
stars \citep{hab96}. The maser emission is detected in
vibrationally-excited, rotational molecular transitions of SiO, denoted
$\nu=\nu',\ J=J'-(J'-1)$, which have rest frequencies of order $\nu_0 \sim
43J'$ GHz. Commonly, the brightest and most frequently observed lines are the
$\nu=1,\ J=1-0$ and $\nu=1,\ J=2-1$ transitions near 43 GHz and 86 GHz
respectively.

In single-dish observations, the SiO maser emission from late-type, evolved stars shows an
appreciable degree of linear polarization: $m_l \sim 15-30\%$, or
greater \citep{tro79}, but a significantly lower degree of circular
polarization, with typical median values $m_c \sim$ several percent
\citep{bar87}.  In a single-dish survey of SiO linear polarization variability in evolved stars, \citet{gle03} report a mean sample linear polarization ${\bar m_l}\sim 23\%,$ with a dispersion of $7\%$. A larger single-dish survey of late-type, evolved stars in full polarization by \citet{her06} was able to classify mean polarization by stellar type or variability class. They report ${\bar{m}_l}\sim 30\%$ and ${\bar m}_c\sim 9\%$ for Mira variables, ${\bar{m}_l}\sim 11\%$ and ${\bar m_c}\sim 5\%$ for semi-regular variables, and ${\bar{m}_l}\sim 5\%$ and ${\bar m_c}\sim 2\%$ for supergiant stars. Overall, the measured single-dish values are consistent with theoretical
expectations, given the non-paramagnetic nature of the SiO molecule
and the associated general theory of the propagation of polarized
maser emission in the limit of small Zeeman splitting \citep{gol73}.

Radio-interferometric observations of SiO maser emission toward
late-type, evolved stars can measure the radiation properties of the
SiO maser emission at high angular resolution in the full set of Stokes
parameters $\{I,Q,U,V\}$. For Very
Long Baseline Interferometry (VLBI), this angular resolution is at the
sub-milliarcsecond (mas) level, and is sufficient to resolve
individual SiO maser components. However, the low level of
mean circular SiO maser polarization, and the
intrinsic millimeter observing wavelengths required for these transitions, pose challenges in
Stokes $V$ instrumental calibration of VLBI observations of SiO
masers. Measurements of SiO maser circular polarization at mas angular
resolution are important however, both to constrain theoretical models
describing the propagation of polarized maser emission, and for the subsequent
application of these
theories to infer the magnitude and orientation of the underlying
magnetic fields from measurements of
component-level SiO maser polarization properties.
Recently, for example,  periodic changes in single-dish fractional circular polarization have been detected toward two SiO stars \citep{wie09}, with a proposed explanation that these changes are tracing precessing planetary magnetospheres in the circumstellar environment \citep{wie09a}.\

 In this paper we consider the instrumental factors that
constrain measurements of SiO circular polarization using VLBI arrays,
evaluate the degree to which these effects can be calibrated and corrected, and
provide an assessment of the net accuracy of SiO circular polarization observations at 43 GHz
and 86 GHz.

The development and first application of VLBI polarimetry was reported by \citet{cot84} and
\citet{war86}. It has since grown to become a mature technique, of broad applicability; see \citet{caw93} and \citet{pol03} for example. For the frequency bands considered in the current paper, the first continuum polarization VLBI were reported at 43 GHz by \citet{kem96}, and at 86 GHz by \citet{att01}. 

These prior works were concerned with
linear polarimetry of continuum extra-galactic radio sources.
\citet{hom99} developed a method to allow circular VLBI polarimetry of continuum sources however, by assuming a source ensemble has $\bar{m}_c\sim0$; this technique has been demonstrated successfully at wavelengths longer than $\lambda \ge 1$ cm \citep{hom09}, and is discussed in further detail below.
The first spectral-line polarimetry in full Stokes$\{I,Q,U,V\}$ at 43 GHz - of the $\nu=1, J=1-0$ SiO maser emission toward the Mira variable, TX Cam - was reported by \citet{kem97}, based on technical development described earlier by  \citet{kem95}. Critical circular spectral-line VLBI polarimetry of water masers in the adjacent 22 GHz band was succesfully performed by \citet{vle02} following this same method.

The current paper examines the sources of instrumental and propagation errors in VLBI circular polarimetry in the $\lambda=7$ mm and $\lambda=3$ mm bands in more detail, and develops and tests refinements to existing calibration methods to improve the accuracy
to which Stokes $V$ and $m_c$ can be measured in such observations. We show that it is possible to refine existing calibration methods based on autocorrelation data that avoid the need for amplitude self-calibration. We demonstrate these refinements on an observed dataset, and show that an error in $m_c$ of $\le 0.5\%$ or better is possible in practice at $\lambda=7$ mm and $m_c \leq 0.5-1\%$ or better at $\lambda=3$ mm. 

The paper is structured as follows: Section 2 describes the instrumental and propagation data model, and Section 3 discusses the associated data reduction methods. The analysis of a test dataset is described in\ Section 4. General discussion and conclusions are presented in Sections 5 and 6 respectively.

\section{Theory}

This section describes the data model, containing propagation and instrumental
effects relevant to precise amplitude calibration of millimeter-wavelength
spectral-line VLBI observations.

Given the focus of the paper described in the Introduction, this discussion
is confined primarily to the case of the VLBA\footnote{The National Radio Astronomy Observatory is a facility of the National Science Foundation operated under cooperative agreement by Associated Universities, Inc.} operating at observing frequencies
of 43 GHz ($\lambda=7$ mm) and 86 GHz ($\lambda=3$ mm). However, these techniques
are broadly applicable to spectral-line VLBI observations at millimeter wavelengths
in general. Implicit to this discussion are the instrumental parameters
applicable to the VLBA; these are summarized in Table~\ref{vlba},
 individually referenced as indicated from  \citet{rom10} or \citet{nap95}. For each of the
two observing bands considered here, the table
contains columns summarizing the receiver band frequency limits, adopted center frequency, typical zenith system equivalent flux density (SEFD), primary beam angular full-width half-maximum $\theta_b$, typical root-mean-square (rms) pointing error $\sigma_p^2$, beam squint separation $\triangle_s$, and the typical baseband bandwidth assigned to an individual maser transition at these observing
frequencies. The beam squint results from the offset Cassegrain optics adopted
in the VLBA antenna design \citep{nap95}. We note also that each of these
VLBA bands is dual polarization, with nominal orthogonal RCP and LCP polarization receptors.


\begin{table*}[]
\caption{VLBA system parameters at millimeter wavelengths\label{vlba}}
\centering
\begin{tabular}{lccccccc}
\colhead{Receiver} & \colhead{Center} & 
\colhead{Zenith} & \colhead{Point source} & \colhead{Primary beam}
& \colhead{Typical pointing} & \colhead{Beam squint} & \colhead{Typical baseband}\\
\colhead{band\tablefootmark{a}} & \colhead{frequency $\nu_0$} & \colhead{SEFD\tablefootmark{a,b}} & \colhead{sensitivity\tablefootmark{a} $P$}
& \colhead{FWHM\tablefootmark{c} $\theta_{b}$} & \colhead{error rms\tablefootmark{d} $\sigma_p$} & \colhead{separation\tablefootmark{e}
$\triangle_s$} & \colhead{bandwidth\tablefootmark{f}}\\
\colhead{(GHz)} & \colhead{(GHz)} & \colhead{(Jy)} & \colhead{(Jy/K)} &
\colhead{(arcmin)} & \colhead{($\theta_{b}$)} & \colhead{($\theta_{b}$)}
& \colhead{(MHz)}\\
\hline
41.0 - 45.0 & 43.1 & 1436 & 13 & 1 & 0.1 & 0.05 & 4,8,16\\
80.0 - 90.0 & 86.2 & 4000 & 40 & 0.5 & 0.3 & 0.05 & 4,8,16\\
\hline
\end{tabular}
\tablefoottext{a}{\citet{rom10}}\\
\tablefoottext{b}{System equivalent flux density - the representative zenith
system temperature multiplied by the point source sensitivity.}\\
\tablefoottext{c}{Full-width half-maximum; $\theta_b=\frac{\lambda}{D}$,for $D$=25 m}\\
\tablefoottext{d}{\citet{nap95} report a typical pointing error rms of 8\arcsec
in 7 ms$^{-1}$ wind conditions.}\\
\tablefoottext{e}{The reported beam squint \citep{nap95} for the offset Cassegrain
optics used in the VLBA\ antenna design.}\\
\tablefoottext{f}{Typical baseband bandwidths used in spectral windows assigned
to individual line transitions at these observing frequencies.}\\
\end{table*}

\subsection{Data Model}

The data model includes all propagation and instrumental calibration effects in the signal path from the frame of the astronomical source to the correlator output. \citet{kem95} present a signal-path analysis of this type for spectral line polarization VLBI in general. We extend
this formalism here to include a more detailed assessment of the problem of amplitude calibration for millimeter observing wavelengths in the case of low, but non-zero, Stokes $V$. For brevity, we use the same notation as the original
paper by \citet{kem95} wherever possible, and enumerate only the data model differences here. These
are presented in the same approximate sequential order
in which they appear in the signal path.

\citet{kem95} presented a reduction heuristic for spectral line polarization
VLBI observations for the case where no prior assumption can be made concerning
the magnitude of Stokes $V$. In contrast for example, in continuum VLBI polarimetry of
compact, extra-galactic sources, the approximation Stokes $V\sim0$ can be
applied during data reduction to a relatively high degree of accuracy. The approach
described by \citet{kem95} defines a data model and develops a full calibration
solution in amplitude, phase, group delay, and fringe rate for a reference
receptor polarization $p \in \{R,L\}$ using standard single-polarization
VLBI techniques;  and then defines a method to tie that solution to the calibration
solution for the orthogonal receptor polarization $q$ by solving for and
applying differential polarization offsets in amplitude, phase, and group
delay. We retain that methodology here, but focus specifically on total and
differential polarization amplitude calibration, and its role in precise
Stokes $V$ measurement at millimeter wavelengths.
This discussion includes a deeper examination
of the data model for autocorrelation data specifically, given their relevance
to amplitude calibration for spectral line VLBI observations as approached here.

\subsection{General}

We carry over several general assumptions made in \citet{kem95}
to the current analysis. Given the small angular size of astrophysical maser
regions observed with VLBI and the millimeter observing wavelengths considered
here, the interferometric image formation problem is in the narrow-field
regime. No correction is required for the non-coplanar baselines effect in
image formation, and all image-plane calibration effects can be treated as
visibility-plane calibration effects. All calibration terms are factorized
accordingly as antenna-based. We adopt a dual-circular orthogonal polarization
basis ($(p,q) \in \{R,L\}$), in keeping with our instrumental focus on the VLBA. 

\subsection{Atmosphere}

Ionospheric Faraday rotation (IFR) produces a differential polarization phase
offset $\gamma^{R-L} \propto \frac{1}{\nu^2}$, where $\nu$ is the observing
frequency. Using a maximum daytime IFR estimated in \citet{tho01} from \citet{eva68} to be
15 rotations at $\nu=100$ MHz for a line-of-sight at a zenith angle $z = 60\degr$, we predict approximate IFR values of $\gamma^{R-L} \sim 0.03\degr$ and $\gamma^{R-L}\sim 0.007\degr$
at 43 GHz and 86 GHz respectively, consistent with intuitive expectations. This effect can therefore be ignored
in the current analysis: $\gamma^{R-L} \ll 1\degr$.

The troposphere is non-dispersive and produces no differential polarization
calibration offsets in any of the calibration quantities. However, the time-variable water vapor and dry atmosphere constituents  are major contributors to absorption in the millimeter-wavelength bands considered
here. We regard the atmosphere as isothermal, with temperature $T_\mathrm{atm}$.
For an atmospheric line of sight of optical depth $\tau$, the noise power
contribution from the isothermal atmosphere is $(1-e^{-\tau})\ T_\mathrm{atm}$.

\subsection{Image-plane effects}

As noted above, the narrow-field constraint applicable in this case (and
in most standard VLBI observations) allows relevant image-plane
effects to be factorized as antenna-based corrections in the visibility plane. This does not detract from the origin of these effects in the image plane however. For the observing wavelengths and specific instrumental emphasis considered here, the most important image-plane effects originate from antenna pointing errors. These pointing errors change the region of the antenna primary beam located toward the angular direction of the correlated field center. This causes
variation in the calibration terms that vary with angular separation from the optical center of the antenna primary beam, such as the amplitude and polarization response.  

The nominal pointing performance and beam squint of the VLBA antennas at
these observing wavelengths are summarized in Table~\ref{vlba}. In the presence of pointing errors, the VLBA primary beam squint ($\triangle_s\sim5\%$ of $\theta_{b}$) introduces a differential polarization amplitude
error. This problem can be analyzed in one dimension without loss of generality. We model the one-dimensional normalized primary beam power pattern at each antenna
along angular
axis $\zeta$ as a Gaussian function of width $\theta_{b}$, beam squint $\triangle_s$,
and pointing error $\epsilon_p$, for nominal orthogonal receptor polarizations $(p,q)\in\{R,L\}$ as:

\begin{eqnarray}
A^p(\zeta) &=& e^{-\frac{4\ln 2}{\theta^2}(\zeta-\frac{\triangle_s}{2}+\epsilon_p)^2}\\
A^q(\zeta) &=& e^{-\frac{4\ln 2}{\theta^2}(\zeta+\frac{\triangle_s}{2}+\epsilon_p)^2}
\end{eqnarray}

The differential polarization amplitude error in this model takes the form:
\begin{eqnarray}
\frac{A^p}{A^q} & = & e^{\frac{8\ln 2}{\theta^2}(\zeta + \epsilon_p) \triangle_s}\nonumber\\
& \simeq & 1 + \frac{8 \ln 2}{\theta^2}(\zeta + \epsilon_p) \triangle_s
\label{eq-pointing}
\end{eqnarray}

\begin{figure}
\includegraphics[width=8cm]{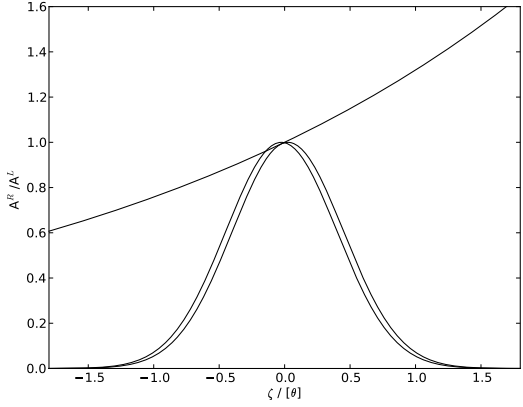}
\caption{A plot of normalized primary beam patterns in 1-d cross-section for receptor polarizations $\{R,L\}$ with a beam squint equal to 5\% of the full-width at half-maximum $\theta_b$; the $x$-ordinate for the beam pattern plot is in units of $\theta_b$. The ratio of the two beam patterns is also plotted in the upper part of the diagram as a function of pointing error $\epsilon_p$, with $x$-ordinate $\frac{\epsilon_p}{\theta_b}$).}
\label{fig-squint}
\end{figure}

The R/L amplitude ratio at the nominal pointing position $\zeta=0$ is plotted as a function of pointing error $\epsilon_p$ in Figure~\ref{fig-squint},
for beam squint $\triangle_s=\frac{\theta_{b}}{20}$, as applicable
to the VLBA. 

This effect is clearly significant for the representative pointing rms values enumerated in Table~\ref{vlba} for the VLBA operating at these observing wavelengths.
We discuss calibration methods for this effect in the second half of this paper.

The net instrumental polarization response at each antenna in each nominal
circular receptor polarization $(p,q) \in \{R,L\}$ can be parametrized
in terms of the net receptor ellipticity and orientation or equivalently
as a complex D-term representing the net leakage of unwanted signal from
the orthogonal nominal polarization \citep{con69}.
The net instrumental polarization response contains contributions from the
antenna optics design, feed illumination function, polarizer performance,
and to a lesser extent, other electronic components in the downstream signal
path that have a differential polarization response. The net instrumental
polarization is an image-plane (direction-dependent) effect but, in the current
narrow-field scientific application, we again factorize it as an antenna-based
visibility plane correction factor, in common with the other image-plane
calibration effects considered in this section. Pointing errors introduce
time-variability in the net instrumental polarization response, expressed
as an antenna-based visibility plane correction, by sampling different regions
of the primary beam polarization pattern, a contributor to the net D-term. The associated dependence on frequency is neglected here due to the small fractional bandwidth.

\subsection{Total power}

Total power relationships for single-dish telescopes are discussed by \citet{kut81}, \citet{dow89},and \citet{wil09}. Adapting this references, we represent the total system temperature at an individual antenna in nominal receptor
polarization $p$  in the form:

\begin{eqnarray}
T_\mathrm{sys}^p &=& \eta_b^p \eta_l T_B^p e^{-\tau} + (1-\eta_l)T_\mathrm{spill} + T_\mathrm{rx}^p \nonumber\\
& & \mbox{} + \eta_l T_\mathrm{atm}(1-e^{-\tau}) \\
T_\mathrm{sys}^p &=&\eta_b^p \eta_l T_B^p e^{-\tau} + T_N^p
\label{tsys}
\end{eqnarray}

where $\eta_b^p=A^p(\epsilon^p)$ is the amplitude loss for an angular pointing
error $\epsilon^p$, the term $T_B^p$ is the source Rayleigh-Jeans brightness temperature
(radiation temperature) in receptor polarization $p$, $\eta_l$ includes antenna scattering and antenna resistive
(ohmic) losses, $\tau=\tau(z)$ is the optical depth along the line of sight at zenith
angle $z$ through the atmosphere, $T_\mathrm{spill}$ is the mean ground spill-over
temperature, $T_\mathrm{rx}^p$ is the net receiver temperature in receptor polarization
$p$, and $T_\mathrm{atm}$ is the mean temperature of the isothermal atmosphere along the line of sight. 

\subsection{Autocorrelation and cross-power spectra}

Autocorrelation and cross-power spectra are formed by interferometric correlation
of all polarization pairs for the voltages recorded in
each nominal receptor polarization at each antenna in the array. We assume
idealized analog correlation here for clarity but without loss of generality - the effects
of quantization, sampling, and other digital signal processing effects are
considered in subsequent sections. Following
the notation of \citet{kem95},  we denote the normalized autocorrelation and
cross-power spectra as $r_{mm}^{pq}$ and $r_{mn}^{pq}$ respectively for antenna indices $(m,n)$ and nominal receptor
polarizations $(p,q)$. The
spectra are normalized by the geometric mean of the total power in the signal
received at each antenna $(m,n)$ forming part of the correlation pair. 

In this Section we express the data model for  $r_{mm}^{pq}$ and $r_{mn}^{pq}$ in the more modern Measurement Equation (ME) matrix formalism introduced by \citet{ham96} and \citet{sau96}.

\subsubsection{Autocorrelation data model}

We denote the normalized analog autocorrelation spectrum as a 4-vector $\mathbb{R}=r^{pq}_{mm}$; this is real-valued for polarization $p=q$ and complex for $p \neq q$.

\begin{equation}
\mathbb{R} = \left( \begin{array}{c}
r^{RR}_{mm} \\
r^{RL}_{mm} \\
r^{LR}_{mm} \\
r^{LL}_{mm} \\
\end{array} \right)(u=0,v=0,\omega)
\end{equation}

where $(u,v)$ are visibility-plane coordinates and $\omega$ denotes angular
frequency. We denote the true source correlation spectrum as $\mathbb{J}$. In the autocorrelation case, this is the (antenna-independent) source total-power spectrum,

\begin{equation}
\mathbb{J} = \left( \begin{array}{c}
J^{RR} \\
J^{RL} \\
J^{LR} \\
J^{LL} \\
\end{array} \right)(u=0,v=0,\omega)
\end{equation}

We denote a total-power noise term contribution $\mathbb{N}^T = (\mathcal{N}^R,\ 0,\ 0,\ \mathcal{N}^L)$, where $\mathcal{N}^p = P^p_m\ T_{N,m}^p$.
Following \citet{kem95}, we use $P^p_m$ to denote the point-source sensitivity
(in Jy/K) of antenna $m$ in receptor polarization $p$. The total system temperature
at antenna $m$ in polarization $p$ is denoted as $T_{\mathrm{sys},m}^p$ and follows equation~\ref{tsys}.

 The autocorrelation data model can then  be represented in matrix form as,

\begin{eqnarray}
&&\mathbb{R}= \mathbb{K}_{mm} \mathbb{L}_{mm}\left(\mathbb{J}+\mathbb{N}_m\right) \label{eq-me}\\
&&\mathbb{K}_{mm} = (\mathbb{G}_m \otimes \mathbb{G}_m^*)(\mathbb{B}_m \otimes \mathbb{B}_m^*)\nonumber\\
&&\mathbb{L}_{mm} =(\mathbb{D}_m \otimes \mathbb{D}_m^*)(\mathbb{P}_m \otimes \mathbb{P}_m^*)\nonumber
\end{eqnarray}

where $\otimes$ denotes an outer matrix product, and $\mathbb{G}_m, \mathbb{B}_m, \mathbb{D}_m$ and $\mathbb{P}_m$ are $2 \times 2$ Jones matrices denoting gain normalization, complex bandpass response, instrumental polarization, and parallactic angle feed rotation respectively. This follows the Measurement Equation formalism of \citet{ham96} but includes a self-noise term, as required in the autocorrelation data model considered here. In this circularly-polarized basis, the Jones matrices take the form, 

\begin{equation}
\mathbb{G}_m=\left(
\begin{array}{cc}
G_m^R & 0 \\
0 & G_m^L
\end{array} \right)
\label{eq-gjones}
\end{equation}

\begin{equation}
\mathbb{B}_m=\left(
\begin{array}{cc}
B_m^R & 0 \\
0 & B_m^L
\end{array} \right)
\label{eq-bjones}
\end{equation}

\begin{equation}
\mathbb{D}_m=\left(
\begin{array}{cc}
1 & D^R_m \\
D^L_m & 1
\end{array} \right)
\end{equation}

\begin{equation}
\mathbb{P}_m=\left(
\begin{array}{cc}
e^{-j\alpha} & 0 \\
0 & e^{j\alpha}
\end{array} \right)
\end{equation}

$\mathbb{K}_{mm}$ is a therefore a diagonal 4 x 4 matrix containing the gain normalization $G_m^p$ and complex bandpass response terms $B_m^p(\omega)$,

\begin{equation}
\mathbb{K}_{mm}=
\begin{array}{c}
{\rm diag} ( \|G_m^R\|^2\ \|B_m^R(\omega)\|^2,\\ 
G_m^R G_m^{L*}B_m^R(\omega)B_m^{L*}(\omega),\\
G_m^L G_m^{R*}B_m^L(\omega)B_m^{R*}(\omega),\\
\|G_m^L\|^2\ \|B_m^L(\omega)\|^2 )
\end{array}\label{eq-gb}
\end{equation}

The net instrumental polarization response $\mathbb{L}_{mm}$ including D-terms and parallactic angle rotation is the $4 \times 4$ matrix,

\begin{equation}
\mathbb{L}_{mm}=\left( 
\begin{array}{cccc}
1 & D^{R*}_me^{-2j\alpha} & D^R_m e^{2j\alpha}& \|D^R_m\|^2 \\
D^{L*}_m & e^{-2j\alpha} & D^R_m D^{L*}_m e^{2j\alpha} & D^R_m \\
D^L_m &  D^L_mD^{R*}_me^{-2j\alpha} & e^{2j\alpha}  & D^{R*}_m\\
\|D^L_m\|^2 &  D^L_me^{-2j\alpha} & D^{L*}_m e^{2j\alpha} &  1
\end{array} \right) 
\label{eq-d}
\end{equation}

\subsubsection{Cross-correlation data model}

The cross-correlation data model does not include a self-noise term; in this case Equation~\ref{eq-me} takes the form,

\begin{eqnarray}
&&\mathbb{R}= \mathbb{X}\ \mathbb{K}_{mn} \mathbb{L}_{mn}\left(\mathbb{J}_{mn}\right) \label{eq-mex}\\
&&\mathbb{K}_{mn} = (\mathbb{G}_m \otimes \mathbb{G}_n^*)(\mathbb{B}_m \otimes \mathbb{B}_n^*)\nonumber\\
&&\mathbb{L}_{mn} =(\mathbb{D}_m \otimes \mathbb{D}_n^*)(\mathbb{P}_m \otimes \mathbb{P}_n^*)\nonumber
\end{eqnarray}

where $\mathbb X$ is a diagonal $4 \times 4$ matrix term for cross-correlation coherence losses; this term does not apply to autocorrelation data. $\mathbb{X}$ does not contribute to differential polarization amplitude offsets, and forms part of the calibration of the absolute flux density scale. 

Here, $\mathbb{J}_{mn}$ is the true source visibility cross-power spectrum,

\begin{equation}
\mathbb{J}_{mn} = \left( \begin{array}{c}
J^{RR}_{mn} \\
J^{RL}_{mn} \\
J^{LR}_{mn} \\
J^{LL}_{mn} \\
\end{array} \right)(u,v,\omega)
\end{equation}

as a function of $(u,v)-$spacing and frequency $\omega$.

\subsection{Sampling and quantization}

The voltages in each nominal receptor polarization are digitally
sampled and quantized by the data acquisition sub-systems at each antenna. Digital realizations $r_{d,mn}^{pq}$ of the total-power and cross-power spectra are obtained by subsequent cross-correlation.
We denote the relationship between the measured spectra $r_{d,mn}^{pq}$ and their true analog counterparts $r_{mn}^{pq}$ used in earlier sections as a general transformation
function $f_d$:

\begin{eqnarray}
f_d:r_{d,mn}^{pq} \to r_{mn}^{pq}\\
f_d^{-1}:r_{mn}^{pq} \to r_{d,mn}^{pq}
\end{eqnarray}

The form of this relationship depends on quantization level, sampler voltage threshold stability and accuracy, and correlator architecture, and is reviewed by \citet{tho01} and references therein. This transformation function does not always take closed analytic form.

 The specific case of sampling and quantization effects for the VLBA correlator is considered by \citet{kog98}. This correlator has an FX architecture, originally hardware-based but more recently upgraded to a VLBA-DiFX software implementation \citep{del07}. Both one- and two-bit sampling are supported. An offset in sampler threshold voltage can produce an associated amplitude error in the output VLBA correlator spectra \citep{kog98}, particularly for two-bit sampling. If the mapping of individual sampler modules to sets of baseband converters mirrors receptor polarization baseband assignment - as occurs naturally for several common baseband converter configurations - these amplitude offsets will translate to differential polarization amplitude errors. The correction of quantization and sampler threshold errors is discussed in further detail in the following section.

\section{Data reduction methods}
\begin{figure}[h!]
\includegraphics[width=7cm]{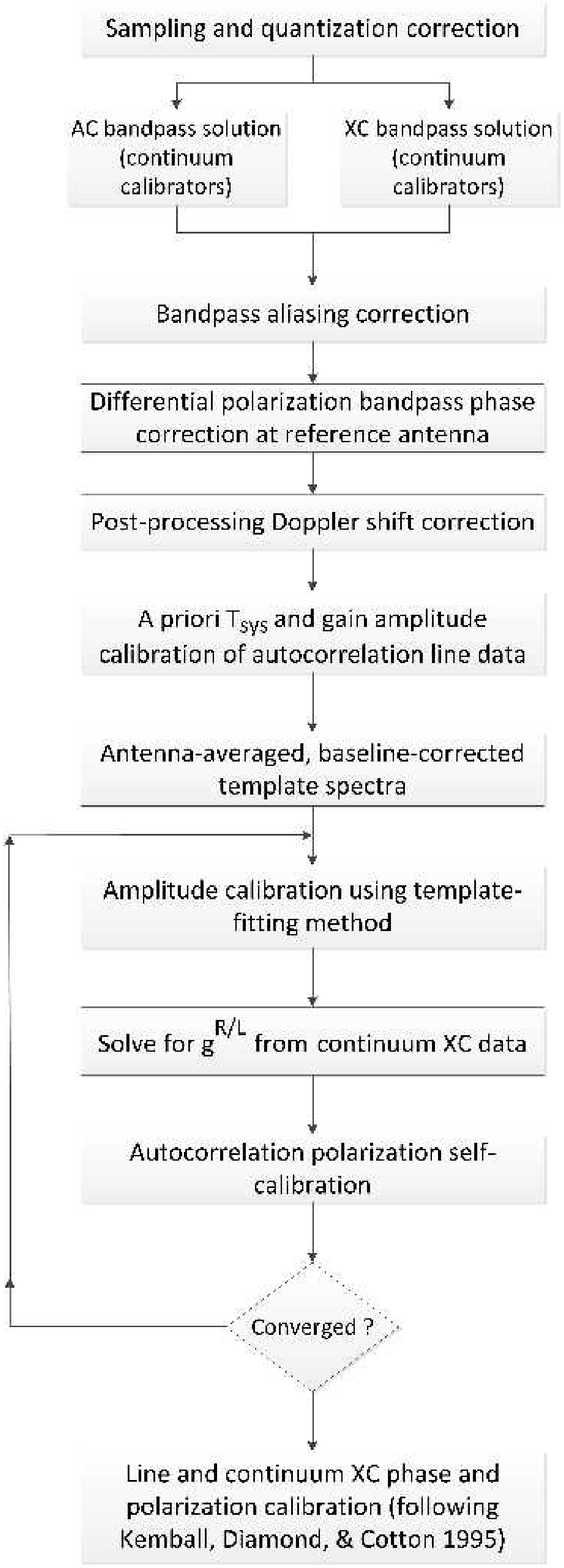}
\caption{A flowchart depicting the overall data reduction sequence described in Section 3.}
\label{fig-flowchart}
\end{figure}

In this section we assess the data reduction methods used to measure and correct the instrumental and propagation terms in the data model enumerated in\ Section 2. Consistent with earlier practice, we describe here only differences with the data reduction methodology presented by \citet{kem95} and use identical notation wherever possible. This section follows the proposed data reduction sequence, which is shown in flowchart form in Figure~\ref{fig-flowchart}.

We have implemented these algorithms within a development framework that brings together several large community codes used for radio-astronomical imaging \citep{kem08}.
For consistency with earlier algorithmic work in \citet{kem95} we have primarily used an adapted implementation of the Astronomical Image Processing System\footnote{http://aips.nrao.edu} within this software framework for the current work. Our intention is to make these revisions publicly available in the future.
 
\subsection{Sampling and quantization}

In this section, we review the correction of digital sampling and quantization effects in VLBI circular polarimetry at millimeter wavelengths. We do not find it necessary to revise existing algorithmic practice, as described in the analysis below.

The total-power and cross-power spectra  $r_{d,mn}^{pq}$ from the VLBA FX correlator are first corrected for quantization effects and then scaled in amplitude by polarization-independent factors known a-priori for the array (the latter known historically as b-factors).
The quantization correction element of the transformation function $f_d$ defined in Section 2.7 above is defined as a relation for each quantization mode (one- or two-bit, $\eta_Q=11$ or $\eta_Q=22$ respectively) between a measured correlation $\rho_m$ and the true underlying correlation $\rho$, both defined in the conjugate Fourier delay lag domain. In lag space, the correlation is of order $O\left(\frac{T_B}{T_B+T_N}\right)$ for cross-power spectra; for the VLBA system temperatures listed in Table~\ref{vlba} and for typical SiO maser spectral flux densities, lag-domain correlation values are small $|\rho_m| \lesssim\ 0.1$. In this case, the relation between $\rho_m$ and $\rho$ is in the highly linear regime \citep{kog98} and the quantization correction can be applied as a direct linear scaling of cross-power spectral values $r_{mn}^{pq}=\alpha_Q\ r_{d,mn}^{pq}$ with numerical coefficient $\alpha_Q$ \citep{kog93a}. As this takes the form of a linear, polarization-independent correction, it cannot produce an artificial instrumental signature mimicking Stokes
$V$. Cross-polarized autocorrelation spectra $r_{mm}^{pq}\ (p\neq q)$ have lag domain correlation values of order  $O\left(\frac{T_B+\sqrt{2}\|D\|T_N}{T_B+T_N}\right)$, as indicated by the first-order form of Equation \ref{eq-me}, and similarly fall in the low-correlation regime. Accordingly they have the same quantization scaling correction as cross-power spectra. 

Parallel-hand autocorrelation spectra $r_{mm}^{pq}\ (p = q)$ have lag-domain correlation values of unity at zero delay, and require Fourier transformation to and from the lag domain to allow application of the full non-linear correction between $\rho_m$ and $\rho$ \citep{kog93a}. As a result, this does not reduce to a linear scaling relationship for the total--power spectra $r_{mm}^{pq}\ (p = q)$ in
the frequency domain. Two effects limit the accuracy of the inverse transform to the lag domain. The VLBA   FX correlator records only positive spectral channels \citep{kog93a}, omitting the opposite sideband which contains digitization noise \citep{tho01}. In addition, an FX correlator requires  equal-length zero padding to  reconstruct the associated correlation function over the full range of sampled delay lags \citep{gra86,tho01}.

Sampler threshold voltages define transition points between discrete quantization states. The non-linear mapping between $\rho_m$ and $\rho$ depends functionally on the value of the sampler threshold voltages relative to their optimal values \citep{kog95}. For two-bit sampling in the low-correlation limit, the relative error in FX spectral output depends linearly on the deviation in threshold voltage level. In one-bit sampling, the relative error in spectral output depends only in second order on the error in the threshold clipping voltage \citep{kog95}. In both VLBA quantization modes, the  correction in output spectral amplitude can be derived from the measured mean deviation of the total-power spectra from the unit mean power level predicted for an ideal digitizer, independent of spectral shape  \citep{kog95}, and we adopt that approach here.

The autocorrelation  data are  used in integrated template-fitting as part of calibration, as described in further detail below. \textit{             }

\subsection{Bandpass calibration}

Following \citet{kem95}, and the Jones matrix formalism in Equation~\ref{eq-bjones} in the current work, we denote the cross-power complex bandpass response as $\bpm$; this is expanded as $\bpm=\|B^p_m\|e^{j\varsigma^p_m(\omega)}$. The autocorrelation bandpass amplitude response, denoted $\|\bpam\|$, differs from the cross-power bandpass amplitude response $\|\bpm\|$ however, due to an unavoidable level of irreducible aliasing in the net system bandpass response. As the continuum calibrators have flat spectra across the baseband bandwidth they are used to solve for $\bpm$ and $\|\bpam\|$ using cross- and total-power continuum calibrator data respectively.

\subsubsection{Bandpass frequency frame}

The VLBA correlator produces output cross- and total-power spectra in a geocentric J2000.0 coordinate reference frame. In contrast, the bandpass response function $\bpm$ is defined in the data acquisition coordinate reference frame  at each antenna, which is an apparent topocentric frame. These two frequency frames are offset by the time-variable natural geometric fringe rate at antenna $m$, denoted $\triangle \omega_m(t)$. The angular frequency of channel number $l$ in the recorded cross- and total-power visibility spectra (which are in a geocentric frame) is denoted as $\omega_l$. The autocorrelation bandpass amplitude response $\bpam$ at antenna $m$ in receptor polarization $p \in \{R,L\}$ can be solved for by minimizing:

\begin{eqnarray}
\left[\chi^{pp}_m\right]^2 & = &\sum_{k=1}^{N_t}\sum_{l=1}^{N_{c}}\left[\|B_{a,m}^p(\omega_l-\triangle \omega_m(t_k))\|^2 - \tilde{V}_{mm}^{pp}\right]^2 \label{bp-ac}\\ 
\tilde{V}_{mm}^{pp} & = & \|V_{mm}^{pp}(\omega_l,t_k)\|\ /\ \overline{\|V_{mm}^{pp}(\omega_l,t_k)\|}\nonumber
\end{eqnarray}

where $V_{mm}^{pp}(\omega_l,t_k)$ is the (real-valued) continuum calibrator total-power spectrum in parallel-hand polarization correlation $pp$ at antenna $m$ over a  pre-average integration interval centered at time $t_k$, the number of frequency channels is $N_c$, and the number of integration intervals is $N_t$. The pre-average interval is fixed and short relative to $\frac{d\triangle\omega}{dt}$. Pre-averaging improves the statistical robustness of the  bandpass response solver, as this is a least-squares minimization problem. The autoscaling of each total-power spectrum by the instantaneous mean spectral amplitude, $\overline{\|V_{mm}^{pp}(\omega_l,t_k)\|}$ , ensures that the spectra do not need to be calibrated in amplitude before solving for $\bpam$. The resulting solution for $\|\bpam\|$ is normalized to unit mean power
$\frac{1}{N_c}\sum_{k=1}^{k=N_c}\|B_{a,m}^p(\omega_k)\|^2 =1$.

Bandpass calibration forms part of visibility amplitude calibration and accurate estimation of Stokes $V$ requires a correspondingly accurate solution for $\|B_m^p(\omega_k)\|$. For example, a spurious bandpass amplitude spike, even if in an off-source spectral region, will bias the mean amplitude normalization of the bandpass over the region of source emission.

The measured visibility spectra and the unknown bandpass response are not in a common frequency frame; the channel shift corresponding to $\triangle \omega_m$ is a non-negligible effect at the wavelengths of $\lambda=3$ mm and $\lambda=7$ mm  considered in this paper. If $\|\bpam\|$ is parametrized discretely as a set of unknown values at each frequency channel (in a stationary topocentric frame) then this can be estimated by direct integration of the total-power, parallel-hand spectra if they are first shifted to the stationary reference frame by using a discrete Fourier transform pair including a channel shift. Alternatively, the unshifted visibility spectra can be fit directly if $\|\bpam\|$ is parametrized as a polynomial series expansion over frequency; this parametrization allows continuous variation of $\triangle \omega_m(t )$ in the formulation of the chi-square minimization problem. This method was developed by \citet{kem97}, using a Chebyshev polynomial expansion  for the bandpass response.
The polynomial bandpass method
avoids digital signal processing artifacts associated with the Fourier transform shift. In addition, as the polynomial expansion order can be significantly lower than the number of frequency channels $N_c$,  this approach reduces the number of free parameters over the discrete parametrization case, thus improving the signal-to-noise ratio (SNR) of the bandpass solution. For these reasons (amongst others discussed below), we adopt the bandpass polynomial solution method. 
In a Chebyshev expansion, the
autocorrelation bandpass response takes the form:

\begin{equation}
\|\bpam\|=\frac{c_{a,0}}{2}+\sum_{j=1}^{N_p}c_{a,j}T_j(x(\omega))
\end{equation}

where $T_j(x)$ is the Chebyshev polynomial of degree $j$
\citep{pre07}, $c_{a,j}$ are real coefficients in the series expansion
for the autocorrelation bandpass amplitude, and the transformed
coordinate is
$x(\omega)=(2\omega-\omega_a-\omega_b)/(\omega_b-\omega_a)$. The
topocentric channel range for the bandpass solution covered by the
measured visibility spectra is $[\omega_a=1-\max(\triangle \omega_m),\
\omega_b=N_c-\min(\triangle\omega_m)]$. In the current work, we
found it important to fit over the full topocentric range
$[\omega_a,\omega_b]$ as opposed to only $[1,N_c]$ as used in
past practice, so that there is the greatest possible chi-square
constraint on the bandpass solution at the edge of the frequency
channel range. The derived solution coefficients $c_{a,j}$ resulting
from the chi-square minimization over $[\omega_a,\omega_b]$ are then
transformed to $c'_{a,j}$ over $[1,N_c]$ as \citep{pre07},

\begin{eqnarray}
&\omega'_k = \frac{\omega_b-\omega_a}{2}\cos\left(\frac{\pi(k-0.5)}{N_c}\right)+\frac{\omega_a+\omega_b}{2}\\
& f'_k = \frac{c_{a,0}}{2}+\sum_{j=1}^{N_p}c_{a,j}T_j(x(\omega'_k)) \\
&c'_{a,j} = \sum_{k=1}^{N_p}f'_k\cos\left(\frac{\pi(j-1)(k-0.5)}{N_c}\right)
\end{eqnarray}

During this coefficient transformation, if $\omega_b<N_c$ or $\omega_a > 1$ then the outer $f'_k$ are extrapolated horizontally to cover the complete range $[1,N_c]$.
 
We note that equation $\ref{bp-ac}$ is formulated for the case of a time-invariant bandpass at each antenna. This is an appropriate instrumental assumption given the expected stability of the net VLBA frequency response, and improves the signal-to-noise ratio of the resulting bandpass solution. In addition a single bandpass solution provides greater control over  the calibration of overall differential polarization R/L amplitude and phase offsets. 

In the most direct formulation, the cross-power bandpass response $\bpm$ can be solved for from the parallel-hand continuum calibrator cross-power spectra $V_{mn}^{pp}$ by an analogous formulation of equation \ref{bp-ac},

\begin{eqnarray}
\left[\chi^{pp}\right]^2 & = & \sum_{m,n>m}^{N_a}\sum_{k=1}^{N_t}\sum_{l=1}^{N_{c}}\left\| B_{m}^p(\omega_{lm})B_{n}^{p*}(\omega_{ln}) -{\tilde{V}^{pp}_{mn}}\right\|^2 \label{bp-xc}\\
\tilde{V}^{pp}_{mn}&=& V_{mn}^{pp}(\omega_l,t_k)\ /\ {\overline{V_{mn}^{pp}(\omega_l,t_k)}}\nonumber\\ 
\omega_{lm} & = & \omega_l-\triangle \omega_m(t_k) \nonumber \\
\omega_{ln} & = & \omega_l-\triangle \omega_n(t_k) \nonumber 
\end{eqnarray}

where $\overline{V_{mn}^{pp}(\omega_l,t_k)}$ is the complex normalization factor to auto-scale the cross-power spectrum in integration $t_k$ over frequency to unit mean amplitude and zero mean phase. 
The auto-scaling applied during the chi-square minimization removes the requirement for amplitude calibration of $V_{mn}^{pp}(\omega_l,t_k)$, as in the case for the autocorrelation bandpass $\bpam$, but for the cross-power bandpass solution, the visibility data need to be calibrated in the phase domain for residual group delay, fringe-rate, phase, and differential polarization phase offset, as described by \citet{kem95}. The resulting solutions for $\bpm$ are normalized over frequency to unit mean power $\frac{1}{N_c}\sum_{k=1}^{N_c}\|B^p_m(\omega_k)\|^2=1$ and zero mean phase.
The polynomial bandpass method is further favored in the cross-power case as there is no unique shift defined for $V_{mn}(t_k)$ if instead of a polynomial bandpass expansion, the visibility data are shifted using a Fourier transform.
\subsubsection{Bandpass aliasing correction}The solutions obtained for the cross-power bandpass response $\bpm$ from the direct formulation in
equation~\ref{bp-xc} are not optimal however for precise amplitude calibration at the millimeter wavelengths ($\lambda$=3 mm and $\lambda$=7 mm) considered here. At these wavelengths, the amplitude of the continuum calibrator cross-power spectra on the longer baselines invariably have insufficient SNR to allow estimation of $\|\bpm\|$ with low mean-square error (MSE) at the outlying antennas. The autocorrelation spectra have significantly higher SNR however, and offer a preferred path for the solution of the cross-power bandpass amplitude response 
 $\|\bpm\|$ if  corrected for aliasing, as found necessary in the current work. We adopt an aliasing model: 

\begin{equation}
\|B^p_{a,m}(\omega_k)\|^2=\|B^p_m(\omega_k)\|^2\ + A_b^2(\omega_{(2N_c-k)})
\label{alias} 
\end{equation}

where $A_b(\omega)$ is a model for the  cross-power bandpass amplitude response at and above the upper end of the band (in a frequency range  $\omega_{k0}>\beta\omega_c$, where $\omega_c$ is the bandpass cutoff frequency and $\beta$ lies in the approximate range $\beta\sim\{0.7-0.9\}$). This function can be extrapolated  and folded to approximate the aliased autocorrelation response.
We find a 12-node Butterworth function \citep{bia07} to give a satisfactory empirical fit to the upper-bandpass region  $\omega_{k0}>\beta\omega_c$, with little sensitivity to the value of  $\beta$ adopted to define the channel range of the fit: 

\begin{equation}
A_b(\omega_k)=a\sqrt{\frac{1}{1+\left(\frac{\omega_k}{\omega_c}\right)^{2n}}}
\label{butterworth}
\end{equation}

 The fit minimizes:

\begin{eqnarray}
[\chi^{pp}]^2&=&\sum_{k=k0}^{N_c}\left[\|B_m^p(\omega_k)\|^2-\left(A_m^p(\omega_k)\right)^2\right]^2\nonumber\\
&&+ \sum_{k=k0}^{N_c-1}[\|B_{a,m}^p(\omega_k)\|^2-\|B_m^p(\omega_k)\|^2\nonumber\\
&&\mbox{}- \left(A_m^p(\omega_{(2N_c-k)})\right)^2]^2\end{eqnarray}

where the number of nodes is fixed at $n=12$, and factor $a$  and the cutoff frequency $\omega_c$ are the only fitted parameters. The direct cross-power solutions for $\|\bpm\|$ obtained from equation~\ref{bp-xc} do however have sufficient SNR to allow a solution for $A_b(\omega)$ in Butterworth form (\ref{butterworth}). The solution for $A_b(\omega)$ allows the high-SNR autocorrelation bandpass amplitude solutions $\|\bpam\|$ to be then transformed to cross-power form:

\begin{equation}
\|B^p_m(\omega_k)\|'=\sqrt{\|B^p_{a,m}(\omega_k)\|^2 - A_b^2(\omega_{(2N_c-k)})}
\end{equation}

In this hybrid approach, the phase $\varsigma_m^p(\omega)$ of the cross-power bandpass response is solved for by minimizing the phase-only analog of equation~\ref{bp-xc}:

\begin{eqnarray}
\left[\chi^{pp}\right]^2 &=&\sum_{m,n>m}^{N_a}\sum_{k=1}^{N_t}\sum_{l=1}^{N_{c}}[ \varsigma_{m}^p(\omega_l-\triangle \omega_m(t_k))\nonumber\\
&& \mbox{} - \varsigma_n^p(\omega_l-\triangle \omega_n(t_k))-\arg(\tilde{V}_{mn}^{pp}(\omega_l,t_k))]^2 
\label{bp-xcp}
\end{eqnarray}

where $\tilde{V}_{mn}^{pp}(\omega_l,t_k)$ are the pre-averaged continuum calibrator cross-power visibility spectra, normalized to unit mean amplitude and zero mean phase, and calibrated in delay, fringe-rate, and phase as described above. In practice, the peak-to-peak residual phase error over frequency for the net VLBA baseband response after correction for phase and group delay is typically of order 5 degrees, and we fit $\varsigma_m^p(\omega)$ at each antenna in each receptor polarization with a low-order polynomial in order to maximize SNR. The complex cross-power bandpass response $\bpm$ is therefore represented by a separate Chebyshev polynomial in each of amplitude and phase, with a lower-order polynomial in phase compared to that in amplitude. The final cross-power bandpass solution is therefore constructed as,

\begin{equation}
\bpm=\|\bpam\|'\ e^{j\varsigma_m^p(\omega)}
\end{equation}

applying the unit-mean power and zero-mean phase normalization to the bandpass solutions as described above.

\subsubsection{Reference antenna differential polarization bandpass phase response}

We note that the solution for $\varsigma_m^p(\omega)$ is derived by
solving a self-calibration problem that is linear in baseline-based
phase, and so is known only relative to the phase at receptor
polarization $p$ for an adopted reference antenna (subscript zero),
i.e. the determined bandpass phase solution takes the form
$\varsigma_m^p(\omega)-\varsigma_0^p(\omega)$, as opposed to
$\varsigma_m^p(\omega)$, the argument of
$B_m^p(\omega)=\|B_m^p(\omega)\|e^{j\varsigma_m^p(\omega)}$ in the
Jones matrix of Equation~\ref{eq-bjones}. A full solution, allowing
correction of cross-polarized visibility spectra therefore requires an
independent estimate of the differential polarization bandpass phase
response at the reference antenna $\varsigma_0^{p-q}(\omega)$. This is
directly analogous to the correction of parallel-hand phase solutions
for differential polarization phase and delay offsets at the reference
antenna, as described by \citet{kem95}. The correction for
$\varsigma_0^{p-q}(\omega)$ has not traditionally been applied to
bandpass corrections in spectral-line VLBI but is relevant to the
autocorrelation polarimetry described in subsequent sections of this
paper, so was implemented in the current work. Without this
correction, for example, there would be no bandpass phase correction
applied to the cross-polarized autocorrelation spectra at the
reference antenna.

\subsubsection{Bandpass correction}

The basic algebra for applying the autocorrelation bandpass response $\bpam$ and $\bpm$ to correct the visibility data is described by \citet{kem95} and \citet{dia89}. These references describe a traditional  $\frac{ON-OFF}{OFF}=\left(\frac{ON}{OFF}-1\right)$ heuristic for autocorrelation bandpass correction, to minimize the residual total-power offset. However, the complete autocorrelation data model used here (Equation~\ref{eq-me}) does not include this term - nor is it necessary - so we do not subtract 1 when correcting the autocorrelation bandpass response in the current work. Additionally, during bandpass application, the fringe-rate shift is applied when computing the polynomial expansion of $\bpam$ or $B_m^p(\omega)$.

\subsection{Amplitude calibration}
Amplitude calibration of the correlation spectra $r_{mn}^{pq}(\omega)$ in units of spectral flux density $J_{mn}^{pq}(\omega)$ (Jy) requires  independent measurement of the total system temperature $T_{\mathrm{sys},m}^p$ (K) and point-source sensitivity $P^p_m$  (Jy K$^{-1}$) both at each antenna $m$  and in each receptor polarization $p$ throughout the course of the observation. We denote the system equivalent flux density (SEFD) as $J_{\mathrm{sys},m}^p=P^p_mT_{\mathrm{sys},m}^p$\ (Jy).
 In the formalism used in this paper (see Equation~\ref{eq-gb} and following), $G_m^p=1/\sqrt{J^p_{\mathrm{sys},m}}$. The $P^p_m$ are
time-variable due to changes in both atmospheric attenuation and antenna gravitational deformation, both as a function of time and antenna pointing position in local horizon coordinates.

Sections~\ref{sec-cont-amp} and \ref{sec-line-amp} contain a review and analysis of current practice in VLBI continuum and spectral-line amplitude calibration, with a specific focus on circular polarimetry at millimeter wavelengths. We describe the innovations in amplitude calibration introduced in the current work in\ Section~\ref{sec-pcal-amp}.

\subsubsection{Continuum amplitude calibration}\label{sec-cont-amp}

The VLBA records integrated system temperatures every two minutes at each antenna in each receptor polarization $p \in \{R,L\}$ obtained using an underlying switched noise calibration system \citep{tho95}. The VLBA project also publishes opacity-corrected gain curves for each antenna, which provide $P^p_m(\tau=0,z)$ as a polynomial function of zenith angle $z$; these curves are obtained from analyses of regular single-dish service observations, separately scheduled \citep{wal99}. 

In standard a priori amplitude calibration of VLBI continuum observations, the $G^p_m(t)$ are computed as $1/{\sqrt{P^p_m(z) T_{\mathrm{sys},m}^p}}$ from the measured system temperature values and published antenna point-source sensitivity curves. At the observing wavelengths $\lambda=7$mm and $\lambda=3$mm considered in this paper the opacity-corrected $P^p_m(\tau=0,z)$ provided by the VLBA require correction for atmospheric attenuation $e^{-\tau}$ (\citet{wal99}; and see Equation~\ref{tsys}). As described by \citet{lep93}, these corrections can be obtained by fitting the measured system temperatures over the course of the observations as a function of zenith angle $z$ against a simplified form of Equation~\ref{tsys}:

\begin{equation}
T_\mathrm{sys}^p=T_a^{p}e^{-\tau}+T_\mathrm{spill}+T_\mathrm{rx}^p+T_\mathrm{atm}(1-e^{-\tau})
\end{equation}

 where $T_a^p$ is the antenna temperature contribution from the source. The number of free parameters can be controlled by adopting a plane-parallel atmosphere, with $\tau=\tau_0 \sec z$ for zenith opacity $\tau_0$, extrapolating $T_\mathrm{atm}$ from ground-level metrology, and adopting an empirical model for the spill-over noise contribution for VLBA antennas \citep{lep93}. In this case, only receiver temperature $T_\mathrm{rx}$ and zenith opacity $\tau_0$ remain as free parameters, and once solved for using chi-square minimization, allow opacity correction in the form:

\begin{eqnarray}
&e^{-\tau(t)} = \frac{T_\mathrm{sys}^p(t)-T_\mathrm{atm}-T_\mathrm{rx}^p-T_\mathrm{spill}}{T_a^p-T_\mathrm{atm}}\\
&P^p_m(\tau,t)=e^{\tau(t)}P^p_m(\tau=0,t)
\end{eqnarray}

The a priori amplitude calibration methods outlined above are limited in their intrinsic accuracy by several sources of error. These include systematic errors in the measured or adopted power level of the noise calibration sources, intrinsic statistical error in the sampled switched $T_\mathrm{sys}$ measurements, and the stability over time of hardware elements affecting amplitude calibration, including receiver gain and the stability of the noise calibration sources.
In addition, the opacity solution is limited by the assumption of a stable, plane-parallel atmosphere; this does not hold at low elevations or if local weather conditions vary significantly over the course of the observations (i.e. $\tau_0=\tau_0(t)$). Other sources of systematic error include the assumption of an isothermal $T_\mathrm{atm}$ extrapolated from the measured ground air temperature, and the relatively coarse empirical model adopted for antenna spill-over noise contributions.

We estimate the overall accuracy of a priori VLBA calibration to be $\sim 10\%$ at $\lambda=7$mm and $\sim 15\%$ at $\lambda=3$mm, excluding low-elevation  data.

A priori amplitude calibration, with or without opacity correction, can be refined using amplitude self-calibration. Historically, two assumptions are often made when performing amplitude self-calibration using parallel-hand visibility data: i) that Stokes $V$  is identically zero; and ii) that amplitude calibration is completely separable from instrumental polarization calibration.  The inapplicability of these assumptions when measuring non-zero fractional circular polarization $m_c  \lesssim$ 0.5\% in compact extragalactic continuum sources is considered in detail by \citet{hom99}, \citet{hom01}, and \citet{hom06}.\ Reduction methods for spectral-line sources with non-zero  $m_c$ are described by \citet{kem95}. As described in the latter reference, in the presence of circular polarization, independent calibration of the parallel-hand correlations $r^{RR}(u,v,\omega)$ and $r^{LL}(u,v,\omega)$ against a total intensity source model that assumes $V=0$ will redistribute circularly-polarized emission in the image, including introducing positional offsets in the centroids of individual circularly-polarized components. As a result, \citet{kem95} used only a single reference receptor polarization $p \in \{R,L\}$ in phase-related self-calibration, and used measured $R-L$ phase offsets to transfer solutions to the orthogonal receptor polarization. Similarly, amplitude self-calibration is avoided, for the reasons noted above. 

\citet{hom99} introduced a method for calibrating continuum sources
with low $m_c$ while still allowing amplitude self-calibration; their
method calibrates $(r^{RR}+r^{LL})/2$ against a common total intensity
source model $I_{mod}$, then solves for long-term residual
differential gain errors at each antenna under the Stokes $V=0$
assumption, i.e. $r^{RR}=I_{mod}$ and $r^{LL}=I_{mod}$.  The latter
measurement of the differential polarization gain factors relies on
the mean circular polarization of the observed ensemble of sources
being close to zero. This approach has been demonstrated successfully
at wavelengths longer than $\lambda\gtrsim 1$ cm
\citep{hom09}. Simulation studies for this technique indicate a
$1-\sigma$ uncertainty in measured circular polarization of $m_c \sim
0.1\%$ for source brightness values exceeding 1 Jy/beam
\citep{hom06}. Uncertainties in amplitude gain calibration predominate
over D-term errors or thermal noise contributions in this method
\citep{hom06}.

The coupling of amplitude and instrumental polarization is evident in the  form\ of Equation~\ref{eq-mex}, which contains both $\mathbb{G}$ and $\mathbb{D}$ terms. Neglecting this coupling is equivalent to linearization of Equation~\ref{eq-mex}, which truncates terms that include instrumental polarization D-terms from the parallel-hand equations \citep{rob94,hom99}.
This can be addressed by an iterative approach to amplitude and polarization calibration \citep{hom99} but with a modest impact on the derived location and magnitude of circularly-polarized components \citep{hom06}.

\subsubsection{Spectral-line amplitude calibration}\label{sec-line-amp}

The template-fitting method derives amplitude calibration from parallel-hand autocorrelation spectra, and was first introduced by \citet{rei80}. The source total power spectrum usually has far greater frequency structure than the continuum noise terms in Equation~\ref{tsys}, and the two can therefore be treated as sufficiently orthogonal in functional form to allow a basis decomposition in terms of the scaled true source spectrum and a residual continuum term varying more slowly with frequency.  The basic method derives time-variable gain normalization factors $G^{p}_m(t)$ by fitting baseline- and bandpass-corrected parallel-hand autocorrelation spectra $r^{pp}_{mm}(\omega)$ to a well-characterized source total power spectrum $J^{pp}(\omega)$ derived from a template scan at a single antenna; this template scan is chosen based on SNR and the quality of available a priori SEFD calibration information for the scan. The method is readily applied to polarization observations by fitting $G^p_m(t)$ separately in each receptor polarization $p$ against the associated parallel-hand template spectra $J^{pp}(\omega)$ and $J^{qq}(\omega)$\citep{kem95}. These authors also introduced a robust method to fit a composite baseline during the template-fit itself, avoiding the need for prior baseline removal in the source and template spectra individually. Given the uncertainties in a priori SEFD calibration discussed above, a secondary correction is needed to determine and remove the differential polarization amplitude gain $g^{R/L}$ tying the template spectra
in each sense of parallel-hand receptor polarization. In the current paper, $g^{R/L}$ is used to refer to a ratio of terms $\|G^R\|$ and $\|G^L\|$  in the Jones matrix of Equation~\ref{eq-gjones}. At centimeter wavelengths, \citet{kem95} solved for this differential polarization gain from relative ratios of the cross-power spectral amplitudes on continuum extragalactic calibrator sources, which were assumed to have Stokes $V\sim0$. We note that in the case of maser sources it is necessary to account for the average antenna noise contribution from the source $\bar{T}^p_{a}=\frac{1}{\delta \omega}\int_0^{\delta \omega}T^p_a(\omega)d\omega$ when transferring the continuum polarization gain ratio to the spectral-line source \citep{kem95}.

The template fitting method has several intrinsic advantages at the wavelengths $\lambda=7$mm and $\lambda=3$mm considered in this paper \citep{kem97}. The method implicitly includes opacity corrections in the determined amplitude scaling factors, and is therefore insensitive to systematic errors in either the opacity parametrization of  Equation~\ref{tsys} or in the published a priori antenna gain curves.
In addition, the method intrinsically tracks short-term variations in the amplitude gain in each receptor polarization at each antenna that result from factors such as pointing errors in the presence of beam squint, or rapid changes in atmospheric attenuation due to highly time-variable local weather conditions. Such short-term amplitude gain fluctuations are significantly more pronounced at wavelengths shorter than $\lambda\leqslant1$ cm. The gain factor information is also encoded in the autocorrelation spectra sampled at the correlator integration rate, as opposed to to a separate sampling interval for switched-noise $T_\mathrm{sys}$ measurement.
The template-fitting method is more statistically robust at measuring short-term amplitude gain fluctuations, such as those caused by pointing errors,  than a method based on detecting associated changes in the measured $T_\mathrm{sys}$ values, as the latter requires separating out changes in two unknown continuum quantities, $\bar{T}_{a}^p$ and the baseline  noise terms. The template-fitting method exploits spectral structure in this decomposition.

\citet{kem97} used two methods
at $\lambda=7$mm to measure the differential polarization amplitude gain $g^{R/L}$ for the template spectra: i) comparison of amplitude self-calibration gain corrections derived from continuum calibrators assuming Stokes $V\sim0,$ obtained for the reference antenna near the time of the template spectrum observation; and, ii) cross-fitting the template spectra in each polarization $J^{pp}(\omega)$ and $J^{qq}(\omega)$ to each other using the template-fitting algorithm to derive the relative differential polarization amplitude gain $g^{R/L}$ directly. Method (ii)  assumes that the small non-zero integrated Stokes $V$ component of SiO emission does not significantly bias the estimated $g^{R/L}$, i.e. that $\bar{m}_c\sim 0$.

In this paper we introduce several refinements to the template-fitting amplitude calibration method to improve its statistical performance in the millimeter wavelength range $\lambda = 3$ mm and $\lambda = 7$ mm considered here. These modifications are described in the following section.

\subsubsection{Autocorrelation polarization self-calibration}\label{sec-pcal-amp}

VLBA system performance degrades sharply toward $\lambda=3$ mm, consistent with array design specifications; this requires that we consider enhancements in the statistical performance of the template-fitting method in the low-SNR regime. In addition, the
high representative fractional linear polarization of SiO masers $m_l \sim 10-30\%$ described earlier suggests that higher-order terms, such as $O(D.(Q+jU))$, that arise in the full non-linear coupling of amplitude and instrumental polarization calibration contained in Equation~\ref{eq-me},
may need to be assessed when using the template-fitting method acting on the parallel-hand autocorrelation spectra.

To explore both concerns, we have implemented an iterative
self-calibration method to derive the amplitude gains
$\mathbb{G}(t)$, instrumental polarization $\mathbb{D}$, and the true
source correlation spectrum $\mathbb{J(\omega)}$ from the measured
un-calibrated autocorrelation spectra $\mathbb{R}(\omega,t)$ at all
antennas.  The initial estimate of the true source spectrum
$\mathbb{J}(\omega)$ is obtained from an weighted average of
baseline-corrected autocorrelation spectra over a restricted range of
low zenith angle and over a subset of antennas with high site
elevation or known low mean precipitable atmospheric water vapor. The
spectra are calibrated a priori by the measured $T_\mathrm{sys}^p$ and
opacity-corrected gain curves $P^p_m(\tau=0,z)$ provided by the VLBA,
and then converted to, and averaged separately in Stokes $\{I,Q,U,V\}$
form. The spectra are pre-averaged at each antenna over a short
interval before baseline subtraction, and weighted by the inverse
mean-square residual error of the baseline polynomial fit. Spectra
with a completeness fraction below a specified fractional threshold of
the pre-average interval are rejected in the global average used as an
estimator for $\mathbb{J}(\omega)$. The averaging of $N$ high-quality
spectra from a subset of antennas reduces statistical noise in the
resulting estimate of $\mathbb{J}(\omega)$ by approximately
$\sqrt{N}$, an important improvement in the low-SNR regime, offset
only by residual systematic error contributed primarily by the
baseline removal process. The baseline is modeled as a low-order
polynomial and removed in a fit to designated off-source spectral
regions. Optionally, auxiliary piece-wise polynomial baselines are fit
and removed above and below the outermost off-source regions; this
approach minimizes the polynomial order required for the primary
spectral baseline. The absolute flux density scale of
$\mathbb{J}(\omega)$ is subject to the uncertainty in a priori VLBA
calibration, estimated above to be $\sim 10-15 \%$ at millimeter
frequencies, and to the approximation of zero atmospheric opacity
$\tau\sim0$ used in deriving the initial source correlation
spectrum. However, our goal here is accurate measurement of the degree
of circular polarization $m_c$; this does not require comparable
accuracy in the absolute flux density scale. For this reason we
enforce the constraint $\int J^{pp}(\omega)\ d\omega = \int
J^{qq}(\omega)\ d\omega$ over the two senses of parallel-hand
correlation in the averaged estimate for $\mathbb{J}(\omega)$, with
corresponding scaling in cross-polarized spectra $J^{pq}(\omega)$. To
establish the global differential polarization amplitude gain
$g^{R/L}$ connecting the mean template spectra $J^{pp}(\omega)$ and
$J^{qq}(\omega)$, we apply the amplitude gains obtained from
template-fitting to these spectra to a compact continuum calibrator
and solve for $g^{R/L}$ by minimizing \citep{kem95}:

\begin{equation}
\sum_{t_k} \sum_{m,n>m} \sum_{\omega_k} \left[\frac{g^{R/L}\|V^{LL\\ }_{mn}(t_k,\omega_k)\|-\|V^{RR}_{mn}(t_k,\omega_k)\|}{g^{R/L}\|V^{LL}_{mn}(t_k,\omega_k)\|+\|V^{RR}_{mn}(t_k,\omega_k)\|}\right]^2\\ \label{eq-uvrat}
\end{equation}

The parallel-hand cross-correlation data $V^{pp}_{mn}(t_k,\omega_k)$ are calibrated in group delay, fringe rate, and by the template-fitted amplitude gains, then pre-averaged
over time and frequency sub-intervals in order to suppress outliers. The statistic in Equation~\ref{eq-uvrat} is by design not sensitive to un-modeled calibrator spatial structure.
Models of intrinsic circular polarization mechanisms in continuum extra-galactic radio sources
\citep{hom09}
suggest intrinsic circular polarization $m_c \ll 0.1\%$ at the millimeter wavelengths considered here.

The parallel-hand auto-correlation spectra are real-valued, unlike the complex cross-polarized autocorrelation spectra, which have non-zero phase and form a conjugate pair $J^{pq}(\omega)=J^{qp*}(\omega)$. Using bandpass solutions corrected for the differential polarization bandpass phase response at the reference antenna \(\varsigma_0^{p-q}(\omega)\)  described earlier, the instrumental phase correction for the autocorrelation cross-polarized spectra (Equation~\ref{eq-gb}) at antenna $m$ can be parametrized in the form,

\begin{eqnarray}
{\rm arg}(\mathbb{K})& = & {\rm diag} (0, \Phi,-\Phi, 0) \\
\Phi & =\ & \phi_m^{R-L} + \omega\tau_m^{R-L} + (\varsigma_m^R-\varsigma_m^L) \label{eq-Phi}
\end{eqnarray}

where $\phi^{R-L}_m$ is the differential polarization phase offset,
$\tau_m^{R-L}$ is the differential polarization group delay, and
$(\varsigma_m^R-\varsigma_m^L)$ is the bandpass phase correction, all
at antenna $m$.  The differential polarization phase offset and group
delay at each antenna are assumed constant over the course of the
observations.  The variation of $\phi_m^{R-L}$ by antenna across the
array introduces amplitude decorrelation when averaging the
autocorrelation spectra to estimate $\mathbb{J}(\omega)$, as described
above. Accordingly, during the averaging process, we normalize the
mean phase (over frequency) of each pre-averaged spectrum to zero
before it is added to the global average estimate of
$\mathbb{J}(\omega)$. This is equivalent to accepting an unknown
additive term in the absolute EVPA of the true source correlation
spectrum $\mathbb{J}(\omega)$; this does not affect our calibration
method, and would traditionally be measured using an astronomical
calibrator of known absolute EVPA.

As evident from Equations~\ref{eq-me} and \ref{eq-d} the amplitude (and for the cross-polarized spectra, phase) of the measured autocorrelation spectra $\mathbb{R}$ change  over both time and frequency due to the combined effect of the polarization leakage terms acting on both the true source correlation spectrum $\mathbb{J}(\omega)$ and the real-valued self-noise $\mathbb{N}$, with the former terms modulated by parallactic angle variation. Our initial estimate of the source correlation spectrum $\mathbb{J}(\omega)$ is accordingly imperfect, as it has been corrected only for an empirical estimate of $\phi_m^{R-L}$ at each antenna, as described above.

Following traditional self-calibration practice, we use our initial
estimate of the source spectrum $\mathbb{J}(\omega)$ to solve for the
instrumental calibration terms. We first hold the initial amplitude
calibration in place, and solve for the instrumental polarization
terms $\mathbb{D}$ at each antenna. This is solved as a non-linear
chi-square problem, minimizing the norm $\|
\mathbb{R}-\left(\mathbb{KL}\right)\left(\mathbb{J}+\mathbb{N}\right)\|$
using full-polarization autocorrelation spectra on the target spectral
line source, summed over pre-averaging intervals $\triangle t$ over
which the parallactic angle does not vary significantly, and over a
spectral channel range encompassing the source emission. This problem
is completely separable for each antenna, as no cross-correlations
between antennas are involved, and is therefore performed separately
and sequentially for each antenna. The free parameters over $p \in
\{\rm{R,L}\}$ are the complex $D^{p}_m$, assumed time- and
frequency-invariant, the differential polarization quantities
$\phi^{R-L}_m$ and $\tau_m^{R-L}$ used to specify $\Phi$
(Equation~\ref{eq-Phi}), and the self-noise contributions
$\mathcal{N}^p$.  The latter are modeled as low-order polynomials over
frequency, in order to accommodate residual errors in bandpass
correction, with independent coefficients per per-average interval due
to time-variable atmospheric terms in Equation~\ref{tsys}. The total
number of free real-valued parameters in the chi-square minimization
per antenna is of order $O(10^2)$ with the greatest contribution from
the independent baseline terms in each pre-average
interval. Accordingly we use two unconstrained minimization methods
suited for the large-parameter case. Given the Gaussian noise in the
measured data, we use the limited-memory BFGS quasi-Newton method
described by \citet{noc80} and \citet{liu89} to find an initial
estimate of the minimum, and a truncated Newton method to refine the
solution \citep{nas84}. Both methods work well in the large-parameter
case.\

Given an estimate for $D^p_m$, $\Phi_m$, and $\mathcal{N}^p(\omega,t)$ obtained from the chi-square solution, we correct the autocorrelation spectra by inverting Equation~\ref{eq-me}, as,

\begin{equation}
\mathbb{J}'=\left(\mathbb{KL}\right)^{-1}\mathbb{R}-\mathbb{N}
\end{equation}

The corrected $\mathbb{J}'=\mathbb{J}_m(\omega,t)$ at each antenna is sampled at the original correlation integration interval and original frequency spacing. For perfect instrumental calibration we would expect an identical estimate of the true source spectrum at each antenna and at all times. However, in common with all self-calibration methodologies we only approach this incrementally over succeeding iterations. Accordingly, we update our estimate of the true source correlation spectrum $\mathbb{J}(\omega)$ by averaging $\mathbb{J}'(\omega)$ over a high-quality subset of data, exactly as described above in connection with the derivation of the  initial estimate.\ Two sub-iterations of polarization calibration are performed before an outer iteration of amplitude calibration using the template-fitting method. In the amplitude calibration step, the corrected $\mathbb{J}'(\omega)$ for the current iteration are fit against the current estimate of the true source correlation spectrum $\mathbb{J}(\omega)$ obtained by averaging followed by an estimate of the differential polarization amplitude gain $g^{R/L}$ as described using Equation ~\ref{eq-uvrat}. This method was found to converge rapidly, generally within one or two outer iterations, to provide a joint estimate for the true source correlation spectrum $\mathbb{J}(\omega)$ and the instrumental parameters $\|G_m^p(t)\|$,\ $D^p_m$, $\Phi_m$, and $\mathcal{N}^p(\omega,t)$.

\subsection{Interferometric phase and polarization calibration}
The cross-correlation phase calibration method used follows that described in \citet{kem95}; namely phase calibration in a reference receptor polarization, combined with the determination of differential polarization phase offsets between the two orthogonal polarization receptors. Interferometric polarization self-calibration is performed as described by \citet{kem97}. Following \citet{hom99}, we add a final phase self-calibration step after interferometric polarization calibration.
 
\section{Results}

In this section we present results obtained by applying the data reduction techniques outlined above to representative observational data, chosen here to be a single epoch from a full-polarization VLBA monitoring campaign of the circumstellar $v=1,\ J=1-0$  SiO maser emission in the $\lambda=7$ mm band toward the Mira variable TX Cam, scheduled as VLBA project code\ BD46. The results of this synoptic imaging campaign have been published in total intensity by \citet{dia03}
and  \citet{gon10}, and in linear polarization by \citet{kem09}. We choose epoch BD46AQ as a representative dataset for this work, as it contains polarization EVPA reversals at the circumstellar boundary and is therefore of independent scientific interest, but for no other special technical reason.
The data were reduced for this epoch following the heuristics outlined in Section 3
and we highlight results relevant to the current data reduction method in this section.

Full details of the observing configuration for  the project BD46AQ are described in the original references above;  we provide only a concise synopsis here.
The effective observing time was 6.5 hours, allocated
between the target source TX Cam and extra-galactic continuum calibrators 3C454.3, J0359+509, and J0609-157.
The $v=1,\ J=1-0$  SiO transition
was centered in a 4 MHz baseband and cross-correlated in full polarization over 128 frequency channels with a correlator accumulation interval of 4.99 s.
The array comprised all ten VLBA antennas and a single antenna from the VLA. The VLBA antenna at Mauna Kea (MK) did not contribute data due to operational difficulties.

\begin{figure}
\includegraphics[width=9cm]{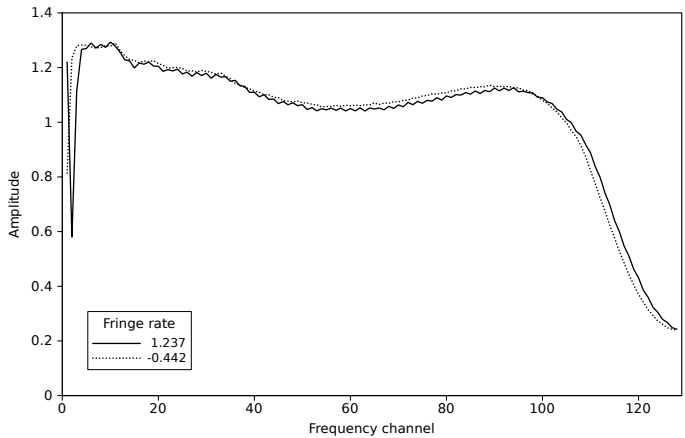}
\caption{Scan-preaveraged parallel-hand (RR) autocorrelation spectra for the two continuum calibrator scans at antenna LA with extremal values of geometric fringe rate over the course of the observing run. Fringe rates are given in units of channel width, and frequency channels 1-128 span the baseband bandwidth of 4 MHz.}
\label{fig-bpgeo-a}
\end{figure}

\begin{figure}
\includegraphics[width=8cm]{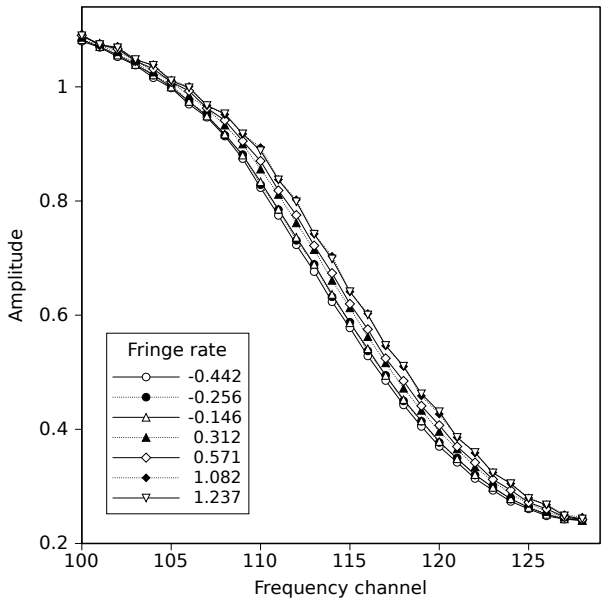}
\caption{Scan-preaveraged parallel-hand (RR) autocorrelation spectra in the upper spectral roll-off region of the baseband for a subset of continuum calibrator scans at antenna LA spanning the range of geometric fringe-rate over the course of the observing run. Fringe rates are given in units of channel width, and frequency channels 1-128 span the baseband bandwidth of 4 MHz. }
\label{fig-bpgeo-b}
\end{figure}

The data were sampled in one-bit quantization; digital corrections for sampling and quantization where derived using methods described above. The derived quantization corrections  have low magnitude $\leq 0.5\%$ and are stable in time, as expected for one-bit quantization. After these corrections, the antenna-based autocorrelation bandpass amplitude responses were solved for in Chebyshev polynomial form using the methods of Section 3.2. The bandpass frequency-frame considerations discussed in that section are illustrated in Figure~\ref{fig-bpgeo-a}; this Figure shows two parallel-hand continuum calibrator autocorrelation spectra produced by the VLBA correlator for a single antenna, here Los Alamos (LA),   pre-averaged
over time in the bandpass solver. These spectra are in a geocentric reference frame, and this Figure shows the two scans that have extremal natural fringe-rate offsets between the topocentric and geocentric frequency reference frames over the course of the observation. As noted in Section 2, the polynomial bandpass solver accommodates the frequency shift algebraically in the Chebyshev expansion for the bandpass response function, without needing to interpolate the sampled data in frequency, and can use fewer free parameters than the total number of frequency channels $N_c$ as a result. The matching full topocentric channel range needs to be used in the fit however, as opposed to $[1..N_c]$, in order to provide a maximal constraint on $\chi^2$ across the band. An expanded view of
the frequency shifts in the scan pre-averaged data near the upper spectral roll-off is shown in Figure~\ref{fig-bpgeo-b}; this depicts a subset of scans spanning the full range of natural fringe rate in the data for the Los Alamos antenna.

\begin{figure*}
\includegraphics[width=16cm]{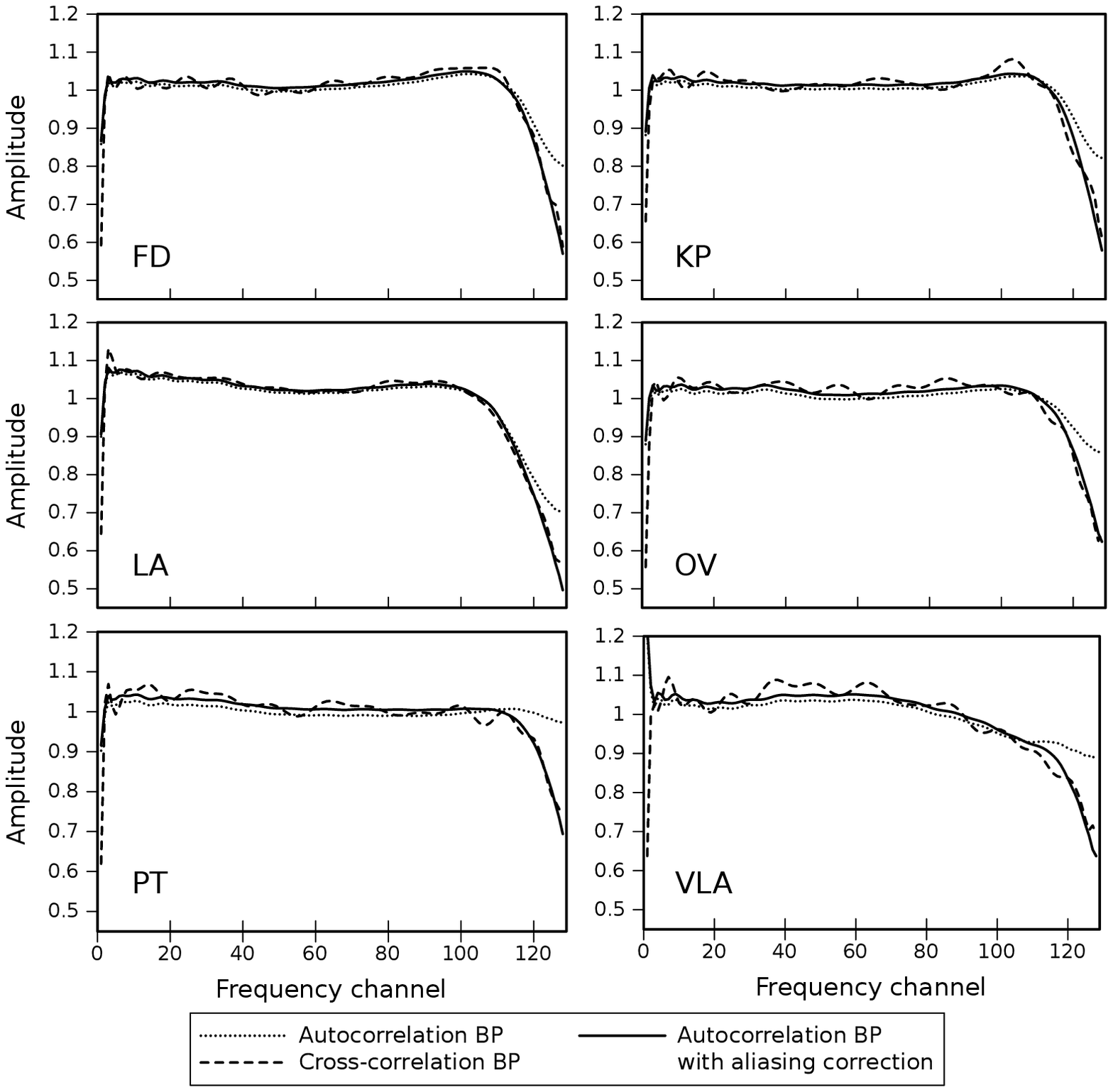}
\caption{Bandpass amplitude response function solutions in RCP for a subset of representative antennas obtained from: i) autocorrelation calibrator data (dotted line) by solving Equation~\ref{bp-ac}; ii) cross-correlation calibrator data (dashed line) by solving Equation~\ref{bp-xc}; and iii) the autocorrelation bandpass solutions corrected for a fitted aliasing model (solid line), as defined by Equation~\ref{alias}}
\label{fig-bpal}
\end{figure*}

As described in Section 3.2, a correction for aliasing is required if the high-SNR autocorrelation bandpass amplitude response functions are to be used to correct the cross-correlation data. The results of the aliasing correction are presented for a selection of representative antennas in Figure~\ref{fig-bpal}; here shown in the RCP bandpass response functions for this antenna subset. For each antenna, a separate amplitude response is plotted for: i) the original autocorrelation bandpass; ii) a cross-correlation bandpass estimate; and, iii) the alias-corrected response. Note the low SNR of the cross-correlation bandpass response obtained from Equation~\ref{bp-xc}, for reasons described in that Section.

\begin{figure}
\includegraphics[width=5.5cm,angle=270]{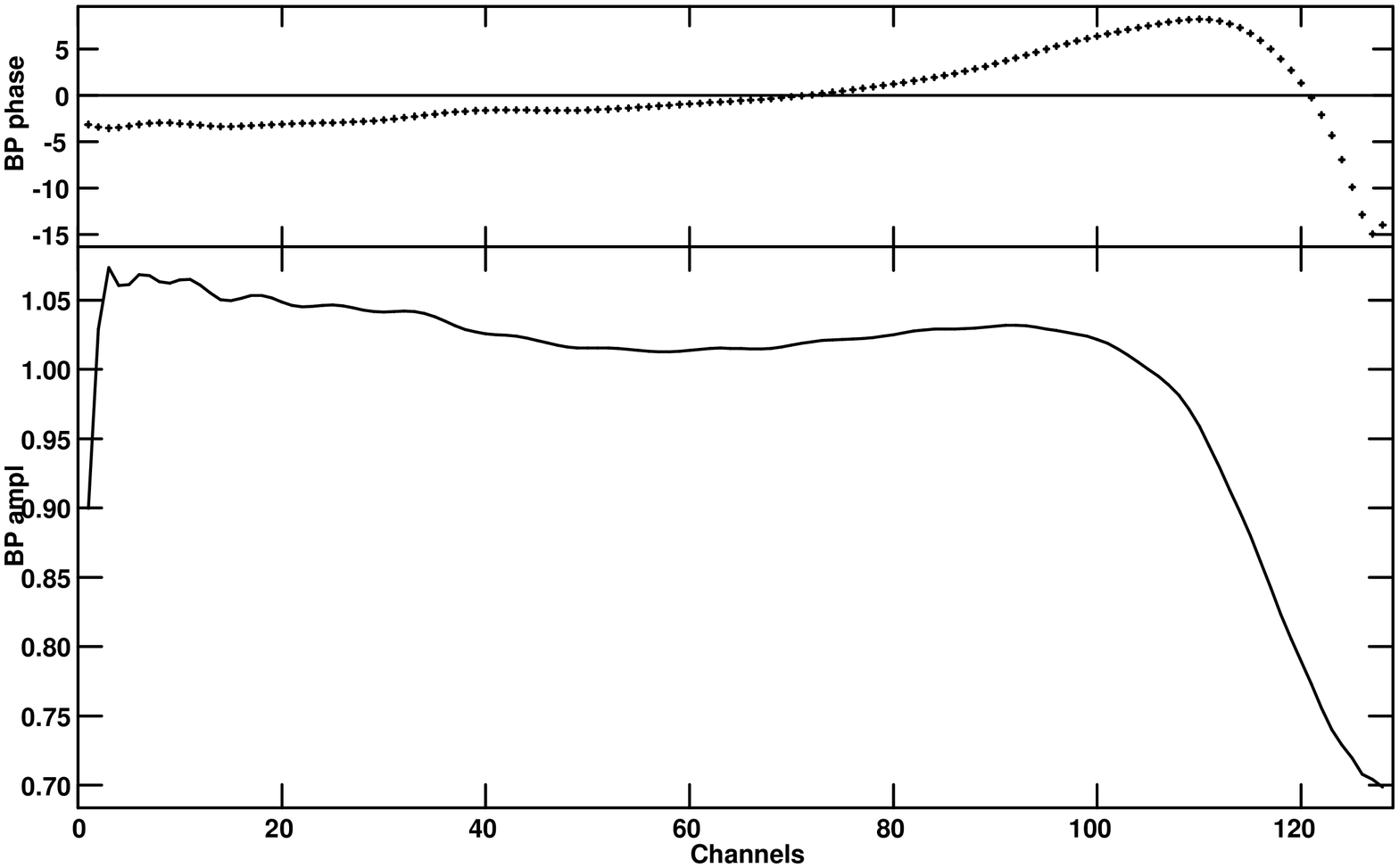}
\caption{The R-L differential polarization bandpass phase response at the reference antenna LA, as measured from the cross-polarized RL and LR autocorrelation spectra. The amplitude response shown here is derived from the parallel-hand autocorrelation data.}
\label{fig-rlphas}
\end{figure}

\begin{figure}
\centering
\includegraphics[width=5.5cm,angle=270]{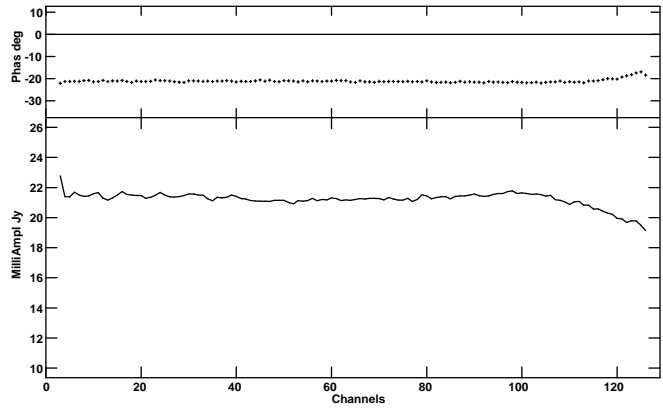}
\caption{The RL autocorrelation spectrum at the reference antenna LA, corrected by the bandpass depicted in Figure~\ref{fig-rlphas}.}
\label{fig-rlphas-b}
\end{figure}

\begin{figure*}
\includegraphics[width=16cm]{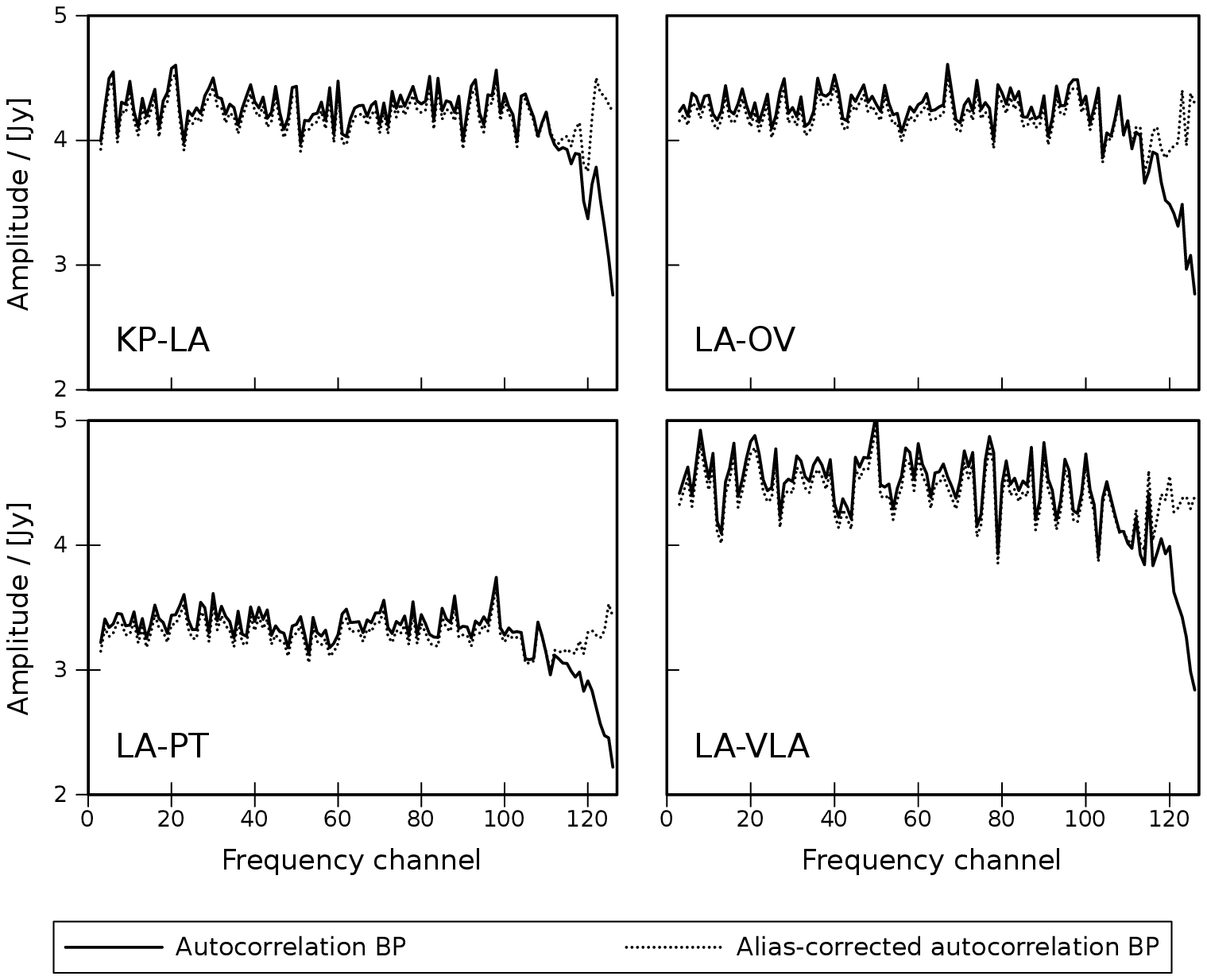}
\caption{Continuum calibrator cross-power spectral amplitudes on representative baselines to the reference antenna LA, calibrated in amplitude, group delay, and fringe rate, and averaged over the duration of the observing run; calibrated additionally either with the original autocorrelation bandpass response (solid line) or the alias-corrected bandpass response (dotted line).}
\label{fig-bpacor}
\end{figure*}

The alias-corrected amplitude response preserves the SNR of the autocorrelation bandpass solution while correcting both the mid-band mean amplitude and upper band-edge frequency response, so that it can be used to correct cross-correlation data more accurately. This improvement in performance when correcting cross-correlation amplitudes is shown in Figure ~\ref{fig-bpacor}.
The final component of bandpass  
determination is the correction for the differential polarization phase response at the reference antenna, derived from the cross-polarized autocorrelation spectra as described in Section 3. . The R-L bandpass phase response as  solved for from the RL and LR autocorrelation spectra at the reference antenna LA is shown in Figure~\ref{fig-rlphas}.
The corrected cross-polarized RL autocorrelation spectrum at LA, after correction for the R-L bandpass phase response is plotted in Figure~\ref{fig-rlphas-b}.

\begin{figure*}
\includegraphics[width=16cm]{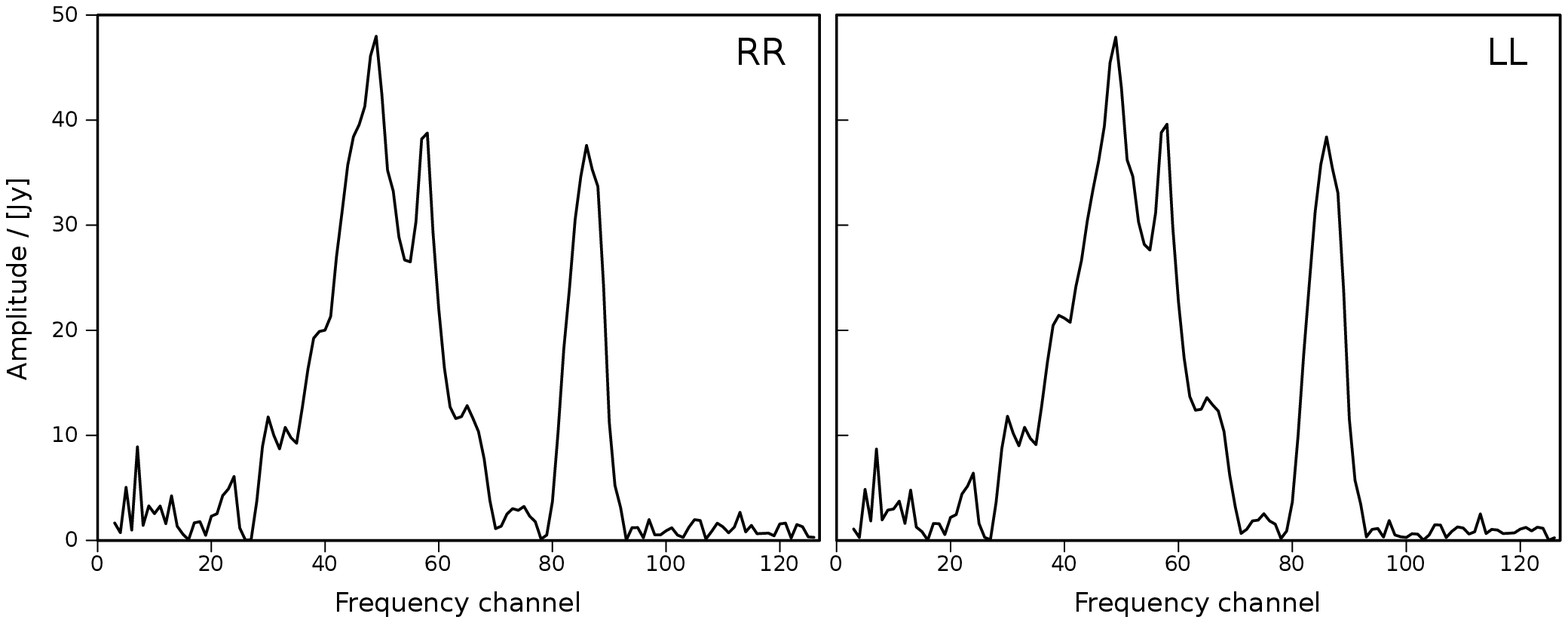}
\caption{Mean autocorrelation template spectra in RR and LL, plotted on a common $y$-axis of spectral flux density (Jy). Frequency channels 1-128 span the 4 MHz baseband bandwidth.}
\label{fig-template}
\end{figure*}

\begin{figure*}
\includegraphics[width=17cm]{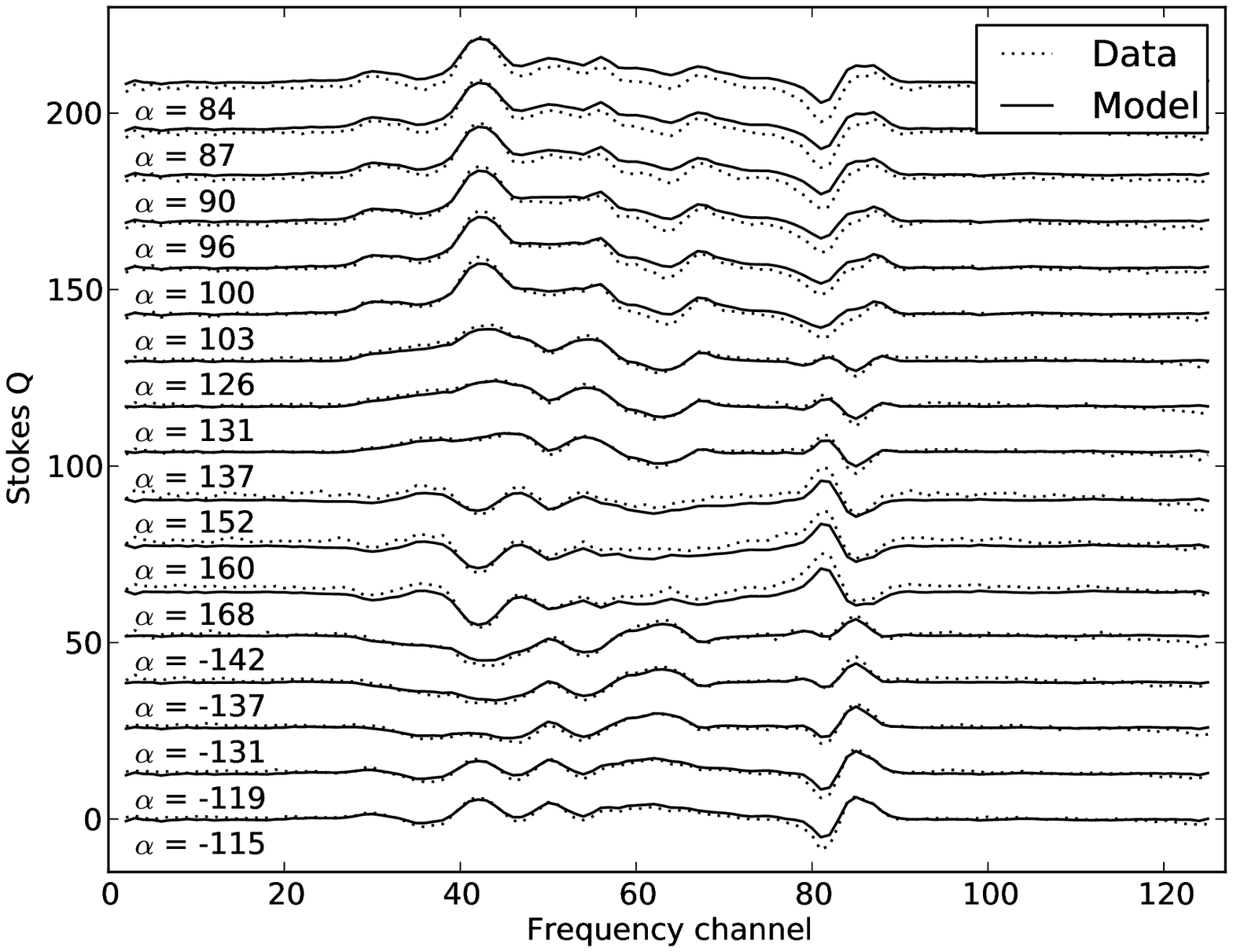}
\caption{Scan pre-averaged autocorrelation Stokes $Q$ spectra before polarization self-calibration (dotted line) and as fitted against Equation~\ref{eq-me} including polarization terms (solid line). Each scan pre-average is labeled at left by parallactic angle $\alpha$ (in deg). }
\label{fig-qufit}
\end{figure*}

\begin{figure}
\includegraphics[width=8cm]{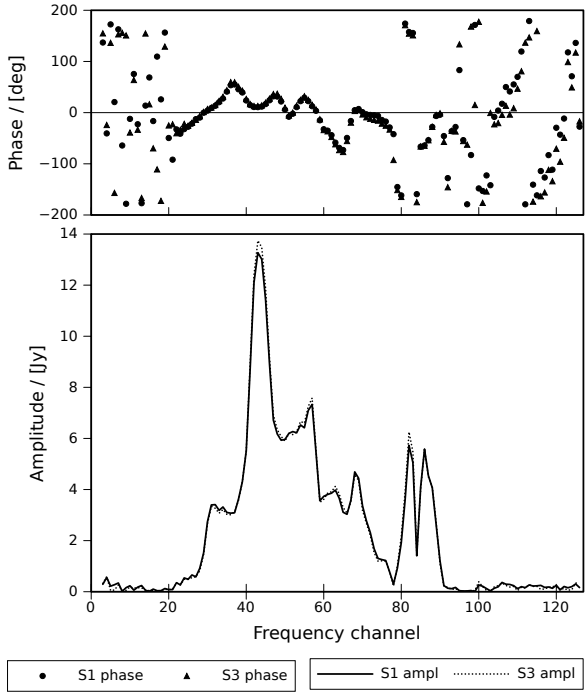}
\caption{Soure RL correlation total-power spectra for iteration one (S1) and three (S3) of the autocorrelation polarization self-calibration algorithm.}
\label{fig-rl}
\end{figure}

The data were then calibrated in amplitude using the autocorrelation polarization self-calibration algorithm described in Section 3.3.3. This method refines the standard template-fitting method to allow the use of a mean multi-antenna template spectrum with higher SNR, depicted here in Figure~\ref{fig-template} for both parallel hands. The method also provides an independent calibration of the differential polarization amplitude gain $g^{R/L}$ (Equation~\ref{eq-uvrat}), and uses the full non-linear autocorrelation polarization data model  including self-noise, as described in Equation~\ref{eq-me}. An example of fitting this equation to the autocorrelation spectra at the reference antenna LA is shown in Figure~\ref{fig-qufit}.
As described earlier, the chi-square minimization in this fit is performed over all Stokes autocorrelation polarization pairs; however for clarity of presentation we show the real-valued Stokes $Q$ spectra pre-averaged over each scan in the form of both the original data and the fitted model. Each scan is annotated by the corresponding  parallactic angle. The time-variability of the Stokes $Q$ spectra across the observation arises
from both the parallactic angle terms in Equation~\ref{eq-me} as well as the D-term mediated corruptions from the autocorrelation spectra in other Stokes polarization pairs. As described in\ Section 3.3.3, an iterative polarization self-calibration method is used to refine both the true source model correlation spectrum and the instrumental terms, including D-terms and the differential polarization phase and group-delay offsets. In Figure~\ref{fig-rl} we show the rapid iterative convergence of the RL source model correlation spectrum over three iterations: the initial and final model spectra are plotted here. 

\begin{figure}
\includegraphics[width=8cm]{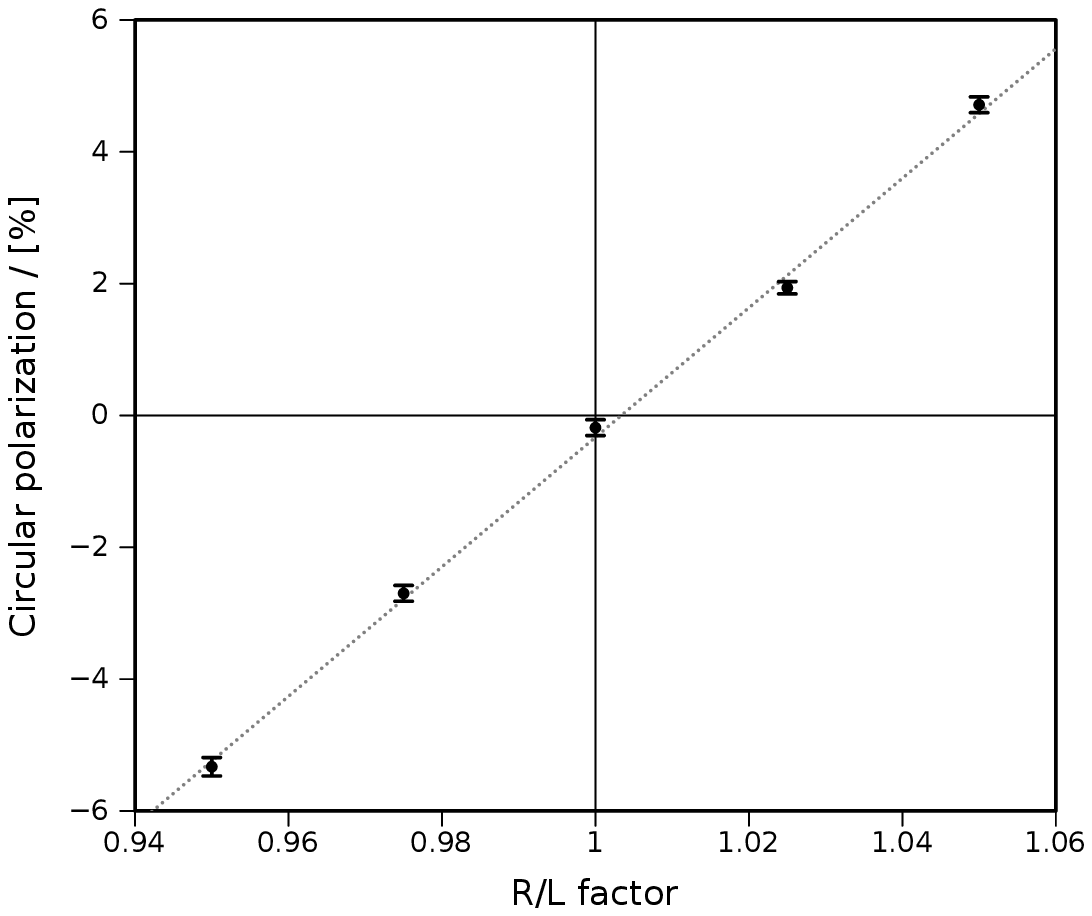}
\caption{Measured circular polarization $m_c$ of the continuum calibrator J0359+509 after application of the TX Cam spectral--line amplitude calibration and  an additional user-specified R/L amplitude gain factor. The continuum calibrator circular polarization measurement for unit additional R/L gain factor provides a single measurement of the error in $m_c$ for the data reduction algorithm presented here. This intercept value is $m_c = -0.3\%$.}
\label{fig-rlfact}
\end{figure}

As a test of net error in Stokes $V$ the amplitude gain factors
derived from TX Cam using the preceding spectral-line calibration
sequence were scaled by an additional constant user-specified R/L
amplitude gain factor and applied to the continuum calibrator
J0359+509. The J0359+509 data were then self-calibrated in phase only,
an independent D-term solution derived using the similarity
polarization approximation \citep{kem99}, and imaged in Stokes
$\{I,Q,U,V\}$. These images were produced over a range of such
user-specified R/L amplitude gain factors. This particular continuum
calibrator was observed most frequently during the schedule, and
amongst the calibrators has the smallest angular separation from TX
Cam on the sky. The integrated degree of circular polarization $m_c$
was then measured by integrating Stokes $V$and Stokes $I$ over a
rectangular region tightly enclosing the central emission in the
calibrator image of J0359+509. The resultant plot of $m_c$ versus the
additional user-specified R/L gain factor is shown in
Figure~\ref{fig-rlfact}.  The measured fractional circular
polarization of the continuum calibrator for unit additional R/L
amplitude gain factor provides an estimate of the error in the data
reduction algorithm presented here. A linear fit produces a value
$m_c=-0.3\%$ at the nominal value where the user-specified R/L factor
is unity.

\section{Discussion}

General-purpose radio-interferometric arrays are optimized in design to provide radiometric accuracy matched primarily to the requirements of their key science goals.  The a priori amplitude calibration information provided by the VLBA, in the form of measured system temperatures $T^p_{\mathrm{sys},m}$ and point source sensitivities $P^p_m(z)$ is not the limiting final accuracy of amplitude calibration; in many cases it can be refined to greater accuracy ($<5\%$) using careful subsequent amplitude self-calibration; and the absolute flux density scale can be set to comparable accuracy using compact amplitude calibrators of known brightness. 

Amplitude self-calibration refinement in this broader context,
traditionally applied to continuum parallel-hand data, requires
imposition of an external constraint on Stokes $V$
however. Invariably, Stokes $V\sim0$ is assumed for individual
continuum calibrator or target sources. However, as noted above, the
more selective constraint that an ensemble of continuum target sources
should have mean ${\bar m_c}=0$ \citep{hom99} allows non-zero $m_c$ to
be measured for individual sources to higher accuracy. For
spectral-line sources, an analogous assumption of zero mean circular
polarization across frequency ${\bar m}_c(\omega)=0$ could be made for
an individual source, however in the latter case this must always be
in essence an ad hoc assumption.  The mean net ${\bar m}_c(\omega)$ for
a given SiO maser source depends on a number of unknown physical
properties of the source, e.g. magnetic field orientation and the
morphological distribution of maser components across the source.

In the current paper we have shown that it is possible to measure
circular polarization in spectral-line VLBI observations at millimeter
wavelengths to a far higher level of accuracy $(< 1\%)$ than suggested
by the absolute accuracy of the a priori amplitude calibration
information $(10-15\%)$, but without the need to perform amplitude
self-calibration. This is possible by relying instead on the amplitude
gain information encoded in the high-SNR autocorrelation spectra,
which are additionally constrained to measure a common
(antenna-independent) source correlation spectrum
$\mathbb{J}(\omega)$. This principle was recognized early in the
development of the template-fitting method of amplitude calibration
for spectral line VLBI \citep{rei80}; in the current work we have
shown that it is possible to enhance this approach sufficiently to
extend it to accurate circular polarization measurements at millimeter
observing wavelengths. This approach has the advantage of avoiding
amplitude self-calibration, which is statistically less robust when
solving for antenna amplitude gain factors across sparse arrays when
observing sources with complex spatial structure.

\begin{figure*}
\includegraphics[width=16cm]{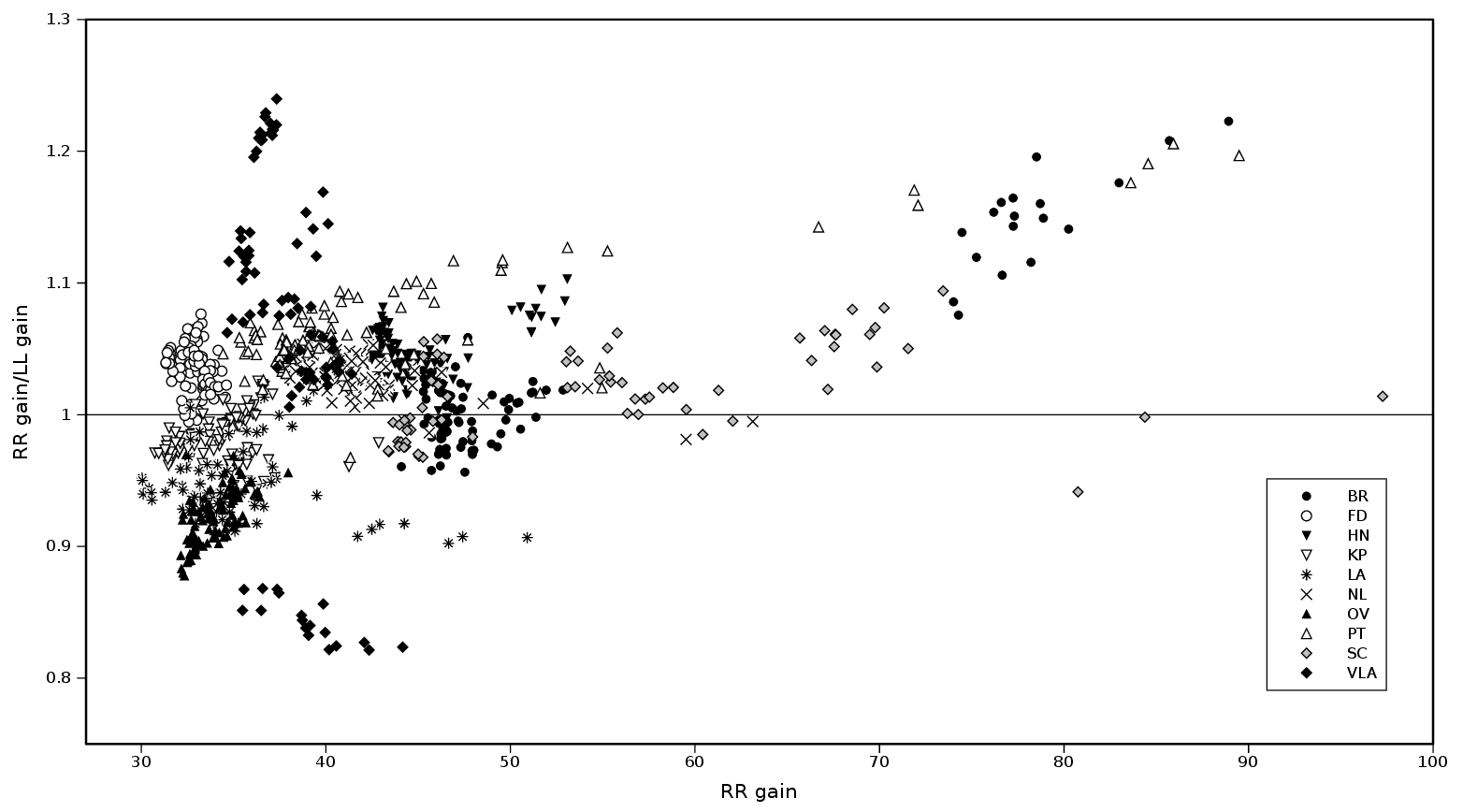}
\caption{Parallel-hand template-fitted amplitude gain factors for each antenna derived from the mean template spectra (Figure~\ref{fig-template}), plotted as differential polarization amplitude gain ratios against the RR amplitude gain derived by the same method.       }
\label{fig-pointing-b}
\end{figure*}

In Figure~\ref{fig-pointing-b}, we show a plot of the polarization
ratio of template-fitted amplitude gains for each antenna in each of
the two orthogonal receptor polarizations against the value of the RR
template-fitted gain derived as part of the same calibration. The
amplitude gain factors are expected to increase with increasing
atmospheric airmass along the line of sight or with increasing
pointing error. These two effects are also partially coupled due to
the expected increase in residual pointing rms at low elevation for
mechanical reasons. The structure in this Figure, especially when
examined for individual antennas, is consistent with differential
polarization amplitude gain effects due to pointing errors in the
presence of beam squint (see Equation~\ref{eq-pointing}), and suggests
the template-fitting method is correcting the associated amplitude
gain effects. This effect is more pronounced at 3 mm, as expected.

We next consider the applicability of the algorithm we have presented in this paper. It does both require sufficient SNR in the autocorrelation spectra, and adequate spectral structure in the source emission so that the source spectrum is sufficiently mathematically non-degenerate relative to continuum baseline terms of polynomial form.
However, these conditions are broadly met in observations of strong maser emission in these bands.

Our implementation of autocorrelation polarization self-calibration here has allowed an assessment the impact of second-order instrumental polarization terms on the parallel-hand autocorrelation spectra, and on the resultant amplitude gain factors estimated using the revised template-fitting method. Our conclusion from the current study is that this correction  is helpful if the autocorrelation data have sufficient SNR and an adequate range of parallactic angle coverage. The SNR condition typically holds at$\lambda=7$ mm but may not be routinely applicable at $\lambda = 3$ mm, due to the significantly higher SEFD in the latter band.
Given the low D-term magnitude
at the VLBA, in the low-SNR case the autocorrelation polarization self-calibration step may not be of net benefit and can be omitted.

Where applicable in terms of SNR however, we find the autocorrelation polarization method to be strongly convergent, especially when deriving an initial estimate of the cross-polarized autocorrelation source model using zero mean phase averaging, as noted above. This convergence is apparent in Figure~\ref{fig-rl}.

We note that the computational cost of the algorithms described in this paper is modest, broadly comparable to regular self-calibration.

Further work would be beneficial in several key areas. An exploration is needed of optimal polynomial bases for the expansion of complex bandpass responses, as a possible alternative to  the Chebyshev basis used here, and especially for the low-SNR regime at these observing wavelengths. In some low-SNR cases the Chebyshev expansion will over-fit the noise and introduce low-level quasi-sinusoidal ripple. Opimized spectral baseline removal methods for the low-SNR case would likely improve both the template-fitting and autocorrelation polarization self-calibration elements.
In addition, and most importantly, the data reduction technique described here needs to be applied to a greater sample of sources  in order to set both the required SNR limit more precisely and to obtain a more comprehensive measurement of the statistical performance of the algorithm as an estimator for $m_c$ and Stokes $V$ over a greater source sample.

From the current work, we estimate the accuracy of the full algorithm presented in this paper as $m_c \leq 0.5\%$ or better at $\lambda=7$ mm, and
$m_c \leq 0.5-1\%$ or better at $\lambda=3$ mm.
We base this estimate
on measurements of Stokes $V$ in associated continuum calibrators, as shown  in Figure~\ref{fig-rlfact} for the test data BD46AQ considered here. These figures are also broadly consistent with sub-sampling bootstrap resampling tests on the variance of $g^{R/L}$ estimated using Equation~\ref{eq-uvrat}.
We note that both these tests tend to over-estimate the error in $m_c$ as both are affected by interpolation errors  from the position of the target source to the position of the continuum calibrator. These interpolation errors include differences in pointing errors and airmass along the lines of sight to the two separate sources.
As noted above, further tests with larger source samples would be helpful in this regard.

The technique presented here however has significant scientific application in the study of circular polarization of astrophysical masers in these observing bands. This is especially true for SiO masers, for which accurate Stokes $V$ measurements provide strong constraints on models of polarized maser propagation, and correspondingly of associated estimates of astrophysical magnetic fields.

\section{Conclusions}
The conclusions of our work are as follows:
\begin{enumerate}

\item{We have examined the sources of calibration error in
spectral-line VLBI imaging observations at $\lambda=7$ mm and
$\lambda=3$ mm sensitive to circular polarization. This analysis was
performed with a specific emphasis on SiO maser observations using the
VLBA.}

\item{A algorithm is presented to provide accurate calibration of
circular polarization without using amplitude self-calibration. This
method is an enhancement of existing spectral-line VLBI calibration
methods based on autocorrelation data, but with several innovations in
bandpass estimation, autocorrelation polarization self-calibration,
and the adaption of techniques for the low-SNR regime applicable at
millimeter wavelengths.}

\item{We demonstrate an example reduction at $\lambda=7$ mm and
provide an estimate of circular polarization accuracy of $m_c \leq
0.5\%$ or better at $\lambda=7$ mm and $m_c \leq 0.5-1\%$ or better at
3mm. These estimates are based on Stokes $V$ imaging of associated
continuum calibrators and a statistical analysis of $uv-$data
differential polarization amplitude ratios.}

\end{enumerate}

\begin{acknowledgements}
We thank the anonymous referee for their insightful comments on this
paper. This material is based upon work partially supported by the
National Science Foundation under grant AST-0507473. Any opinions,
findings, and conclusions or recommendations expressed in this
material are those of the authors and do not necessarily reflect the
views of the National Science Foundation.
\end{acknowledgements}

------------------
\end{document}